\documentclass[11pt]{article}
\usepackage{geometry} 
\usepackage{graphicx}
\usepackage{amssymb}
\usepackage{rotating}
\newcommand{\cO}{{\cal O}}
\begin{document}
\title{\hfill\textbf{\small CERN-PH-TH/2010-131}\\\hfill\textbf{\small OUTP-10-14P}\\
Beyond MFV in family symmetry theories of fermion masses}
\author{ Zygmunt Lalak, Stefan Pokorski \footnote{Hans Fischer Senior Fellow, Institute for Advanced Studies, Technical University, Munich, Germany} \\
{\normalsize \textit{Physics Department, University of Warsaw,} }\\
{\normalsize \textit{Institute of Theoretical Physics
Ul. Hoza 69
PL-00 681 Warsaw, Poland}%
}\\\\
Graham G Ross \\
{\normalsize \textit{CERN, 1211 Geneva 23 } }\\
{\normalsize \textit{and} }\\
{\normalsize \textit{Rudolf Peierls Centre for Theoretical Physics,} }\\
{\normalsize \textit{University of Oxford, 1 Keble Road, Oxford, OX1 3NP, UK}%
}}
\date{}
\maketitle

\abstract{ Minimal Flavour Violation (MFV)  postulates that the only source of flavour changing neutral currents  and CP violation, as in the Standard Model, is the CKM matrix. However it does 
not address the origin of fermion masses and mixing and models that do
usually have a structure that goes well beyond the MFV framework.
In this paper we compare the MFV predictions with those obtained in models based on spontaneously broken (horizontal) family symmetries, both Abelian and non-Abelian. The generic suppression of flavour changing processes in these models turns out to be 
weaker than in the MFV hypothesis. Despite this, in the supersymmetric case, the suppression may still be consistent with a solution to the hierarchy problem, with masses of superpartners below $1$ TeV.  A comparison of FCNC and CP violation in processes involving a variety of different family quantum numbers should be able to distinguish between various family symmetry models
and models satisfying  the MFV hypothesis. }

\section{Introduction}
The Standard Model (SM) provides an accurate description of all the presently available experimental data for flavour changing neutral current (FCNC) and CP violating processes.
Their precision is already good enough to leave room only for small corrections from the physics beyond the SM (BSM). Thus, if the scale of new physics is O(1TeV), as is relevant for the hierachy problem, its flavour structure must be strongly constrained.
An interesting hypothesis that is consistent with the present constraints is that the physics beyond the Standard Model satisfies the principle of Minimal Flavour Violation (MFV): according to it the only source of FCNC and CP violation, as in the SM, is the CKM matrix. 
The MFV conjecture can be implemented in some concrete BSM theories. For instance, it is satisfied in the MSSM with universal soft scalar masses and coefficients of the trilinear soft terms proportional to the associated Yukawa couplings. The new FCNC and CP violating effects are then small enough to be consistent with the data even for squark masses well below 1TeV. However if the MSSM is extended to include a spontaneously broken family symmetry MFV is violated even if, before spontaneous family symmetry breaking, the soft scalar masses are universal and the coefficients of the trilinear soft terms are proportional to the associated Yukawa couplings.

The phenomenological implications of the MFV hypothesis can be investigated in an elegant model independent way by using an effective field theory approach (EFT) \cite{D'Ambrosio:2002ex}. In this framework the SM lagrangian is supplemented by all higher dimension operators consistent with the MFV hypothesis, built using the Yukawa couplings treated as spurion fields. The potential deviations of the data from the SM predictions are then parametrized in terms of few free parameters such as
the inverse (messenger) mass scale associated with the higher dimension operators, with their flavour structure fixed by the structure of the CKM matrices. 


The MFV hypothesis relies on the phenomenological knowledge of the CKM matrix and implicitly assumes that the eventual theory of fermion masses is consistent with it. However, this may not be the case. Indeed, explicit theories of fermion masses and mixing usually violate the MFV hypothesis and it is the purpose of the present paper to investigate this problem. Our laboratory will be Froggatt-Nielsen-like models, \cite{Froggatt:1978nt}, \cite{Leurer:1992wg}, \cite{Ibanez:1994ig}, \cite{Dudas:1995mm},  \cite{Binetruy:1996xk}, \cite{Irges:1998ax},  \cite{King:2001uz},
with spontaneously broken family symmetries and familon field(s) whose vacuum expectation values (vevs) determine the Yukawa couplings
(for an earlier discussion on the possible violation of the MFV hypothesis in models with broken family symmetries see 
\cite{Feldmann:2006jk, Hiller:2010dv} and
\cite{Altmannshofer:2009ne}). In \cite{Altmannshofer:2009ne} a detailed phenomenological analysis has been performed for the MSSM with some Abelian and non-Abelian \cite{King:2001uz} family symmetries. Following \cite{D'Ambrosio:2002ex} we will analyse this case using the SM EFT approach. Horizontal symmetries must then be imposed on the higher dimension operators of the effective SM and the familon fields can be used in their construction as spurion fields. 

Although the effective field theory approach is quite general, care must be taken when interpreting the bounds on the messenger mass scale because the interpretation does depend on the nature of the new physics. This occurs if there is more than one scale associated with BSM physics. We shall illustrate this problem with a detailed discussion of the SUSY case in which there are two basic scales, the SUSY breaking scale and the family messenger scale. In this case it is useful to apply the EFT approach above the SUSY breaking scale in the manner suggested in \cite{Gabbiani:1988rb}, and we extend our family symmetry analysis to cover this approach too.

We first review the MFV hypothesis for the SM viewed as an EFT. We then construct the analogous higher dimension operators in Froggatt-Nielsen like theories using the spurion technique generalized to this case (for an earlier discussion of the use of the spurion technique beyond MFV see \cite{D'Ambrosio:2002ex} and in models with family symmetries see \cite{Feldmann:2006jk}). We illustrate the expectation by comparing the bounds on the effective messenger scale obtained in MFV and in a variety of family symmetry models that have been proposed to explain the observed pattern of fermion masses and mixings. In the second part of the paper we discuss the problem of the interpretation of the effective messenger scale in supersymmetric models and extend our analysis to an EFT description above the SUSY breaking scale.
In this paper we will consider only flavour changing processes originating in the quark sector.

\section{Minimal Flavour Violation and beyond}

\subsection{MFV}

The SM fermions consist of three families with two $SU(2)_L$ doublets ($Q_L$ and $L_L$ ) and 
three $SU(2)_L$ singlets ($U_R$ , $D_R$ and $E_R$ ). Each of these fields is a triplet in flavour space. The largest group of unitary field transformations 
that commutes with the gauge group is $U(3)^5$. This can be decomposed as 
\begin{equation}
G_F = SU(3)^{3}_q \otimes SU(3)^{2}_l \otimes U(1)^5, 
\end{equation}
where 
$SU(3)^{3}_q = SU(3)_{Q_L} \otimes SU(3)_{U_ R} \otimes SU(3)_{D_R}$, $SU(3)^{2}_l = SU(3)_{L_L} \otimes SU(3)_{E_R}$. 
The symmetry is broken by the Yukawa interactions, 
\begin{equation}
\mathcal{ L} = \bar{Q}_L Y_D D_R H 
+ \bar{Q}_L Y_U U_R H_c 
+ \bar{L}_L Y_E E_R H + h.c.,
\end{equation}
where $H_c= i \tau^2 H^*$ and $<H^\dagger H > = v^2 /2$. Treating the Yukawa coupling matrix as spurion fields transforming as
\begin{equation}
Y_U \sim (3, \bar{3}, 1)_{SU(3)^{3}_q}, \; Y_D \sim (3, 1, \bar{3})_{SU(3)^{3}_q}, \; Y_E \sim (3, \bar{3})_{SU(3)^{2}_l} 
\end{equation}
the full Lagrangian has an $SU(3)^5$ invariant form.

MFV postulates that the only source of $G_F$ breaking are the Yukawa spurions and parameterises the higher dimension flavour violating terms by using them to construct the most general $SU(3)^5$ invariant set of higher dimension operators that make up the full effective field theory. The leading terms are the dimension 6 operators given in Table 1. Following \cite{D'Ambrosio:2002ex} it is convenient to write them in terms of products of two-fermion operators which separately should be $SU(3)^5$ invariant, because the flavour structure of all the operators of Table 1 is determined by the flavour structure of these two fermion operators. In particular the four fermion operators factorise into the product of two fermion operators. As we shall discuss this factorisation does not always apply beyond MFV.

Due to the smallness of the down quark Yukawa couplings the dominant operators displayed in Table 1 have external down quarks for which the up Yukawa couplings are responsible for the flavour changing terms. The leading two-fermion operators from which one may determine the MFV predictions for the operators of Table 1 are \begin{equation}
\bar Q_LY_uY^{\dagger}_uQ_L, \, \bar D_R Y^{\dagger}_d Y_uY^{\dagger}_u Q_L .
\label{dim3MFVops}
\end{equation} 

The flavour structure of these operators is determined by the flavour structure of Yukawa matrices. In the electroweak basis where the down-type quarks are mass eigenstates (EWDD), to a very good approximation, it is determined by the entries proportional to the \textquotedblleft square" of the top quark Yukawa coupling: $(Y_uY^{\dagger}_{u})_{ij}$ where $i,j$ are flavour indices. In this frame 
$ \lambda_{FC}= (Y_uY^{\dagger}_{u})_{ij}\approx \lambda_t^2 U^{*}_{3i}U_{3j}$
where the matrix $U$ is the CKM matrix.
The relative magnitude of various FCNC effects is determined by the order of magnitude of the mixing angles and their absolute values depend in addition on the ratios of the couplings of those operators over the (unknown) scale of new physics that has been integrated out.

For the sake of easy reference, in Table 1 we quote the bounds on the suppression scale $\Lambda$ from ref. \cite{D'Ambrosio:2002ex}, obtained by using the measured values of the mixing angles. Here, the scale $\Lambda$ is defined as an effective scale, with the operator coupling equal to 1. If the new physics contributes  e.g. only at the loop level,
the bound on its actual physical scale is lower by factor $\alpha$.

%
\renewcommand\arraystretch{1.2}
\begin{table}[h]
$$
\begin{array}{rc|cc}
\multicolumn{2}{c}{\hbox{Flavour violating}} &\multicolumn{2}{c}{\Lambda_{MFV} }(in TeV)\\
\multicolumn{2}{c}{\hbox{dimension~six~operator}} &
- & + \\
\hline
\cO_{0}= &\frac{1}{2} (\bar Q_L \lambda_{\rm FC} \gamma_{\mu} Q_L)^2 
\phantom{X^{X^X}}
&6.4 & 5.0  \\
\cO_{F1}= & H^\dagger \left( {\bar D}_R \lambda_d \lambda_{\rm FC} \sigma_{\mu\nu}
Q_L \right) F_{\mu\nu}  &9.3 & 12.4  \\
\cO_{G1}= & H^\dagger \left( {\bar D}_R \lambda_d \lambda_{\rm FC} \sigma_{\mu\nu}
T^a Q_L \right) G^a_{\mu\nu}  & 2.6 & 3.5 \\
\cO_{\ell1}=& (\bar Q_L \lambda_{\rm FC}\gamma_{\mu} Q_L)(\bar L_L \gamma_\mu L_L )  
& 3.1 & 2.7 \\
\cO_{\ell2}= &( {\bar Q}_L \lambda_{\rm FC} \gamma_\mu \tau^a Q_L)
({\bar L}_L \gamma_\mu \tau^a L_L)\quad  & 3.4 & 3.0 \\
\cO_{H1}=& (\bar Q_L \lambda_{\rm FC} \gamma_{\mu} Q_L)(H^\dagger i D_\mu H)\qquad  
& 1.6 & 1.6  \\
\cO_{q5}=& (\bar Q_L \lambda_{\rm FC} \gamma_{\mu} Q_L)(\bar D_R \gamma_\mu D_R )  
& \multicolumn{2}{c}{\sim 1} \\
\end{array}$$
\label{table1}
\caption[X]{{\bf
Bounds on the suppression scale of the dimension 6 operators in the MFV scenario. }
{\em The SM is extended by adding flavour-violating
dimension-six operators with coefficient $\pm1/\Lambda^2_{MFV}$
($+$ or $-$ denote their constructive or destructive
interference with the SM amplitude). D'Ambrosio et al. \cite{D'Ambrosio:2002ex}
report the bounds at $99\%$ {\rm CL} on $\Lambda_{MFV}$, in TeV,
for the single operator (in the most representative cases).}
}
\end{table}

\subsection{The Messenger scale}

The dimension 6 operators of Table 1 appear in the effective Lagrangian multiplied by a factor $1/\Lambda^2$ that has the dimension of two inverse powers of mass. This factor arises due to the propagator of the messenger state that is responsible for generating the operator and that has been integrated out when constructing the effective Lagrangian relevant at energy scales less than the messenger mass. In phenomenological studies the lower limit on this factor is determined and gives an estimate of the possible scale of new physics. However some care is needed in interpreting this limit because there may be more than one messenger scale involved. 
In particular, in a realistic extension of the Standard Model there usually exists a mechanism easing the hierarchy problem, with an associated mass scale $\Lambda_h$. This role could be played by supersymmetry with the characteristic scale of mass splittings in supermultiplets, $M_{SUSY}$, or by a strongly coupled gauge theory with the confinement scale $\Lambda_{conf}$  or by the mass scale of Kaluza-Klein states in Randall-Sundrum models. The sector responsible for the flavour violation has its own characteristic scale, which we shall call the family messenger scale $M$  which can be larger than $\Lambda_h$ or coincide with it. 
The effective Lagrangian is relevant at energy scales less than the messenger mass $M$ and less than $\Lambda_h$. Depending on the details of the theory, the suppression factor could be one of the following: $1/\Lambda_{h}^{2}, \ 1/ (\Lambda_{h} M), \ 1/M^2$.  If $M \gg \Lambda_h$, operators suppressed by only the first factor will be the most important. We will return to a detailed discussion of the identification of $\Lambda$ in supersymmetric models in Section 5.

\subsection{Beyond MFV}
MFV is based on the very restrictive assumption that the Yukawa couplings  are the only source of flavour symmetry breaking. This assumption is not valid for many (most) of the attempts to build a theory of fermion masses and mixing and so it is of interest to develop a formalism capable of describing such models and highlighting the main discrepancies to be expected from MFV.

Consider the case of the two fermion operators just discussed. The most general set of nontrivial $SU(3)^3_q$ representations of the two fermion operators that can be made up of quarks and antiquarks is
\begin{equation}
(3, \bar{3}, 1), \;(\bar{3}, 3,1), \; (3, 1, \bar{3}), \; (\bar{3}, 1,3), \;  (1, 3, \bar{3}),  \;  (1, \bar{3},3),\;(8, 1,1),\;  (1, 8,1),\;  (1, 1,8)
\end{equation}
In MFV, c.f. equation (\ref{dim3MFVops}), the fundamental Yukawa couplings transform as $(\bar{3},3,1)$ and $(\bar{3},1,3)$ and these must be combined with the quark bilinears to form $SU(3)^3_q$ invariants corresponding to  the dimension 6 four fermion operators of Table 1. However in models of fermion mass there may be spurions,  combinations of fundamental familon fields with non-vanishing vacuum expectation values (vevs), with different $SU(3)^3_q$ transformation properties to those of the Yukawa couplings. This then leads to new possibilities for the construction of four quark operators. For example in reference \cite{Feldmann:2006jk},  the effect of fundamental spurions transforming as $(8,1,1)$ was studied in detail. However, as stressed below, building $SU(3)^3_q$ invariant combinations of four quark operators and familon fields typically involve familon combinations, i.e. spurions,  transforming in  {\it all} possible
$SU(3)^3_q$ representations, not necessarily with correlated magnitude, in a manner that does not correspond to building four fermion operators  starting from a single fundamental spurion. 

An important consequence of this is that family symmetries often require fewer insertions of the familon fields than would be expected in MFV. For example to construct the (8+1,1,1) representation in MFV requires two Yukawa spurion insertions, $\bar Q_LY_uY^{\dagger}_uQ_L$ involving LR and RL couplings  at the messenger level but can be directly constructed from familon fields in a manner not involving the RH sector.

\section{Family symmetry models.}
In this paper we shall be concerned with the departures from MFV to be expected in models of fermion masses and mixings based on spontaneously broken family symmetries. A wide variety of family symmetries have been considered, varying from one or more Abelian family symmetries or their discrete subgroups to non-Abelian symmetries or discrete non-Abelian symmetries. Such models have been shown to be able to generate the hierarchical structure of quark masses and mixing angles. To illustrate the implications for FCNC we will consider a variety of representative models.

The first two models \cite{Chankowski:2005qp} have a single Abelian family symmetry factor and a single familon field whose vacuum expectation value (vev) spontaneously breaks the symmetry. The third (supersymmetric) model 
\cite{Ivo:2006ma} also has a single Abelian factor but has two familon fields that acquire equal vevs along a D-flat direction. In addition the Higgs field can carry a charge under the symmetry. The model generates a texture zero that leads to a precise prediction for the Cabbibo angle in excellent agreement with experiment. The fourth model \cite{Leurer:1993gy} involves two Abelian factors. Unlike all the other models considered here, in the current quark basis, the dominant off-diagonal term generating the Cabibbo angle comes from the up- and not the down-quark mass matrix. The fifth model involves a Non-Abelian family symmetry and the model was developed to describe both quark, charged lepton and neutrino masses and mixing. The group is a discrete non-Abelian subgroup of $SU(3)$ family symmetry, the discrete subgroup chosen because it leads to near tri-bi-maximal neutrino mixing in agreement with experimental measurements. However the structure of the low dimension terms is determined by the $SU(3)$ symmetry and so for the discussion here it does not matter that only a discrete subgroup is unbroken. Finally we consider a model \cite{King:2010mq} with three Abelian factors  based on the structure found in F-theory string models \cite{Dudas:2009hu} in which the family symmetry is a subgroup of the underlying $E(8)$ string symmetry. In this the emergence of three Abelian factors is natural and unlike the previous models the charges of the fermions are strongly constrained by the $E(8)$ symmetry. 

 For the case that the symmetry is Abelian, all the independent $SU(3)^3_q$ representations of spurions bilinear in the fermion fields are generated at a fundamental level. A subset of the dimension 6 four fermion operators are also fundamental and cannot be built from the two fermion operators, i.e. they do not factorise. As we shall discuss this leads to a potential enhancement of flavour violation.  For the case the symmetry is non-Abelian, as for MFV only a restricted set of fundamental spurion representations bilinear in the fermion fields are present and the dimension 6 operators may be built from them.
 
 One may worry about the possible effects of Goldstone modes resulting from the spontaneous breaking of the family symmetry. For the case that the family symmetry is a local gauge symmetry the familons provide the longtitudinal component of the family gauge boson. If the symmetry breaking scale is large these bosons will not appear in the effective low energy lagrangian and their effect will be negligible. For the case the family symmetry is a discrete symmetry there are no Goldstone modes and the familons can be very heavy. In what follows we will not consider the case that the family symmetry is global and so we will not discuss thepossible effects of  massless familons.
 
We start with a discussion of the quark bilinear operators relevant to the structure of quark masses and to the construction of higher dimension operators. In the next Section we extend the analysis to the dimension 6 operators relevant to flavour changing processes.
The set of dimension 3 operators that violate flavour are given in Table \ref{deftab1a}. In this we have suppressed the family index so, for example,  $\bar{Q}_L X^{Q}_{LL} Q_L = \bar{Q}_L ^iX^{Q}_{LL,ij} Q_{L,j}$ for $i,j=1,2,3$. 

\begin{table}[htdp]

\begin{center}
$$
\begin{array}{|c|c|}
\hline \hline 
1. & \bar{Q}_L X^{Q}_{LL} Q_L \\
2. & \bar{D}_R X^{D}_{RR} D_R \\
3. & \bar{U}_R X^{U}_{RR} U_R \\
4.& \bar Q_L X^{D}_{LR}{D}_R \\
5.& \bar Q_L X^{U}_{LR} {U}_R\\
\hline \hline
\end{array}
$$
\end{center}

\caption{\em{\bf Flavour changing dimension 3 operators in the Standard Model.} The associated  Lorentz and colour structure is not shown.}

\label{deftab1a}
\end{table}%

As discussed above, for the case of MFV only the first and the fourth operators are needed to construct the dimension 6 flavour changing operators, the remaining ones give negligible contributions due to the smallness of the down quark Yukawa couplings. However for family symmetries all operators can be significant.  We turn now to a discussion of the magnitude of the coefficients, $X$, of these operators.

\subsection{Abelian family symmetry}
Consider a $U(1)$ family symmetry. Up to coefficients of order unity the elements of the Yukawa matrices are given in terms of the family charges of fermions defined as 
$q_i$ for the flavour components of the left-handed doublet $Q_L$, and $u_i$ and $d_i$ for the 
flavour components of the (left-handed) quark singlet fields $U^c$ and $D^c$, the charge conjugate of the right-handed flavour triplets $U_R$ and $D_R$, respectively. 

We first consider the holomorphic case in which the symmetry is spontaneously broken via familons carrying only one sign of $U(1)$ charge. For a single familon, $\theta$, with $U(1)$ charge equal $+1$ the Yukawa matrix of couplings has the form
(the $U(1)$ charge of the Higgs doublet is taken to be zero) \footnote{We work in the canonical basis for the kinetic terms.
The rotation from a non-canonical basis to the canonical one does not change our considerations, see \cite{Leurer:1993gy}, \cite{King:2004tx}.}
\begin{equation}
\bar{Q}_L Y_U U_R H_c = \bar{Q}_{L} ^i  \left [a_{i} ^j\left (\frac{\theta}{M} \right )^{u_j+q_i} \right ] U_{Rj} H_c \;~~{\rm  if} \;u_j+q_i\ge 0,\; {\rm otherwise}\;=0,
\label{holomorphic}
\end{equation}
where $a_{i}^j$ are coefficients of order unity and $\theta$ now denotes the familon vev. Note that this structure applies to the superpotential (F-terms) in supersymmetric theories because supersymmetry does not allow terms involving the conjugate of the chiral superfields. Non-supersymmetric theories do not have this restriction so for them the non-holomorphic form discussed below applies. The same is true of D-terms in supersymmetric theories.

Given this we turn to the structure in  the non-holomorphic case. In supersymmetric theories this applies to F-terms too for the case there are familon fields with the same charge but  of both sign. This is very common in supersymmetric models where the family symmetry breaking familon fields $\theta$ and $\bar{\theta}$ with $U(1)$ charges +1 and -1 acquire equal vevs along a D-flat direction. We are denoting this common vev by $\theta$. As just mentioned the non-holomorphic case also applies to the D-terms and to non-supersymmetric theories because in them the symmetries allow terms involving the familon or its conjugate. In all these cases the Yukawa couplings take the form 
\begin{equation}
\bar{Q}_L Y_U U_R H_c = \bar{Q}_{L} ^i\left [a_{i} ^j\left (\frac{\theta}{M} \right )^{|u_j+q_i|} \right ] U_{Rj} H_c 
\label{nonholomorphic}
\end{equation}
To avoid unnecessary duplication of formula we will use the notation ${|u_j+q_i|}$ to denote both the cases of equations (\ref{holomorphic}) and equation  (\ref{nonholomorphic}). In practice the holomorphic form is only relevant to the form of the fermion mass matrix in SUSY theories; the non-holomorphic form applies to the operator coefficients in all cases.

We assign to the combination $a_{i}^j\left ( \frac{\theta}{M_m} \right )^{ |u_j +q_i|}$ transformation rule as for $(3,\bar{3},1)$ under $SU(3)^{3}_q$.  One can regard the 3x3 matrix of the coefficients $a^{j}_i$ as a spurion field transforming as
$(3,\bar{3},1)$ under $SU(3)^{3}_q$ and the factors $\Phi_{L}^i=(\theta/M)^{q_i}$ and $\Phi^{\dagger \, i}_{u}=(\theta/M)^{u_i}$
as  $U(1)$ spurions which are singlets under the flavour group.\footnote{We thank A.Weiler for a useful discussion of this
point.}
It is notationally convenient to write this as 
\begin{equation}
a_{i}^j \left (\frac{\theta}{M} \right )^{ |u_j+q_i|}= a_{i}^j\Phi_{L}^i\Phi_{Uj}^{\dagger }\equiv \Phi_L a^{LU}\Phi^{\dagger}_U
\end{equation}
where $\Phi_{L}$=$((\theta/M)^{q_1}$,$(\theta/M)^{q_2}$,$(\theta/M)^{q_3})$, ($\Phi^{\dagger}_u=(\theta/M)^{u_1}$,....) and the modulus in the exponent is to be taken for the combined charges. 

In terms of the familons the quark Yukawa lagrangian reads
\begin{equation}
\mathcal{ L}_Y = ( \bar Q_L \Phi_La^{LD}\Phi^{\dagger}_D D_R) H
+ (\bar Q_L \Phi_La^{LU}\Phi^{\dagger}_U U_R) H^c +h.c.
\label{Yuk}
\end{equation}
The quark bilinears in these terms correspond to the operators 4 and 5 of Table 2. The remaining operators can be constructed in an analogous way, with the help of familons and horizontal and flavour symmetries giving:
\begin{equation} \label{dim6flavons}
\bar Q_L \Phi_La^{LL} \Phi^{\dagger}_L Q_L,~~ \, \bar U_R \Phi_Ua^{UU} \Phi_U^{\dagger} U_R,~~ \, \bar D_R \Phi_D a^{DD}\Phi_D^{\dagger} D_R.
\end{equation}
where the matrices of $O(1)$ coefficients $a^{IJ}$ are not related  and transform as $(8,1,1)$, $(1,8,1)$ and $(1,1,8)$, respectively for $I,J=LL,UU,DD$.

The above analysis is readily extended to the case that the family symmetry is $U(1)_L \times U(1)_R$. In this case $\Phi_{L}$=$((\theta_L/M_L)^{q_1^L}$,$(\theta_L/M_L)^{q_2^L}$,$(\theta_L/M_L)^{q_3^L})$ and 
\newline
$\Phi_{R}$=$((\theta_R/M_R)^{q_1^R}$,$(\theta_R/M_R)^{q_2^R}$,$(\theta_R/M_R)^{q_3^R})$ where we have allowed for different messenger scales associated with the familon fields breaking the left and right $U(1)$ symmetries. 

\subsection{Non-Abelian family symmetry}
There has been a proliferation of models based on non-Abelian symmetries driven by the possibility that they can explain the near bi-tri-maximal mixing observed in the lepton sector through neutrino oscillation experiments. It is only through a non-Abelian structure that Yukawa couplings to different families can be related {\it including} the $O(1)$ coefficients and this is needed to generate bi-tri-maximal mixing. Although the motivation comes from the lepton sector it is natural to try to extend the symmetry to include quarks and for this reason we include a discussion of non-Abelian family symmetries here. 
Again the family symmetry must be chosen to be a subgroup of $SU(3)^3_q$. Here we consider the simple case that the non-Abelian family group is the diagonal $SU(3)$ subgroup or a discrete subgroup of it. The symmetry is broken by familon fields in a definite representation of the symmetry. For the case that the LH and charge conjugate RH fields are in the triplet representation they acquire a vev the form $\Phi=(c_1,c_2,c_3)$ where $c_i$ are constants. This field must be used to build the Yukawa couplings and the higher dimension operators. Thus the quark Lagrangian may contain terms of the form
\begin{equation}
\mathcal{ L}_Y = \alpha_D \bar Q_L \Phi\Phi^{\dagger} D_R H/M^2_D
+\alpha_U \bar Q_L \Phi \Phi^{\dagger} U_R H^c /M^2_U+h.c.,
\end{equation}
where we have allowed for different messenger masses in the down and the up sector. The parameters $\alpha_{U,D}$ are (family independent) constants and the relative magnitude of the Yukawa matrix elements is set by the constants in $\Phi$. In practice, in order to generate the observed masses and mixing angles, several familon fields are necessary. The remaining two quark operators are constructed in a similar manner and have the form 
\begin{equation}
\alpha_{LL}\bar Q_L \Phi \Phi^{\dagger} Q_L/M_D^2, \, \alpha_{UU}\bar U_R \Phi \Phi^{\dagger} U_R/M_U^2, \, \alpha_{DD}\bar D_R \Phi \Phi^{\dagger} D_R/M_D^2.
\label{naops}
\end{equation}
In what follows we will compare the prediction of MFV with four representative family models. The structure of these models is given in Appendix 1.

\section{Comparison of MFV and family symmetry models - Dimension 3 quark bilinear operators}

In this Section we compare the predictions for the magnitude of the FCNC effects based on the MFV conjecture with those
to be generically expected in models with family symmetries. 
For the case the operators of Table 2 involve down quarks it is necessary to work in the electroweak basis with diagonal down quark Yukawa matrix (EWDD), as used in the beginning of this section to discuss the MFV results. 

For the case the operators of Table 2 involving up quarks it is necessary to transform to the basis in which the up quarks are diagonal before estimating the coefficients. As mentioned above, these operators are negligible in the MFV case due to the smallness of the down Yukawa couplings but may be significant in the family symmetry case.

The mass eigenstate (primed) basis is obtained by rotating right and left fields, 
\begin{equation}
D^{'}_R = V_{D}^{\dagger} D_R, \; U^{'}_R = V_{U}^{\dagger} U_R, \;\bar{Q}_L = \bar{Q}^{'}_L  S^{\dagger}_d .
\end{equation}
where $S_d,\;V_{U,D}$ are unitary matrices. In the EWDD basis the down Yukawa couplings are diagonal so the Lagrangian has the form 
\begin{equation}
\mathcal{ L} = \bar{Q}^{'}_L Y_{Dd} D^{'}_R H
+ \bar{Q}^{'}_L S^{\dagger}_d S_u Y_{Ud} U^{'}_R H_c + h.c.,
\end{equation}
where the subscript $d$ denotes diagonal matrices. The CKM matrix is $U = S^{\dagger}_u S_d$.
From this point on we work in the EWDD basis but drop the primes. 
To determine the matrices $S_{u,d}$ we must diagonalise the associated mass matrices. Following from equation (\ref{Yuk}) the up and down mass matrices have the form
\begin{equation}
M^u_{ij}\propto \left(\frac{\theta}{M_U}\right)^{ |q_i+u_j| },\;\;M^D_{ij}\propto \left(\frac{\theta}{M_D}\right)^{ |q_i+d_j|} ,
\end{equation}
where we have allowed for different expansion parameters in the up and down sectors. The first two and the fourth Abelian family symmetry examples presented in Appendix 1 have the same expansion parameter in the up and down sectors, $M_U=M_D$. The third Abelian example and the non-Abelian example both allow for different expansion parameters.

We write the mass matrices in the form

\[
M=m_{3}\left( 
\begin{array}{ccc}
\widetilde{m}_{1} & \epsilon _{1} & \epsilon _{2} \\ 
\epsilon _{1}^{\prime } & \widetilde{m}_{2} & \epsilon _{3} \\ 
\epsilon _{2}^{\prime } & \epsilon _{3}^{\prime } & 1%
\end{array}%
\right) .
\] 
For the models considered here the matrix can be written in leading order of powers of $\epsilon=\theta/M$ and up to coefficients of $O(1)$ as
\[
M=m_{3} \left (
\begin{array}{ccc}
1 &\epsilon_1/\tilde{m}_2&\epsilon_2\\
-\epsilon_1/\tilde{m}_2& 1&\epsilon_3 \\
- \epsilon_2&- \epsilon_3&1
\end{array}
\right ) 
\left (
\begin{array}{ccc}
\tilde{m}_1&0&0\\
0& \widetilde{m}_2&0 \\
0&0&1
\end{array}
\right ) 
\left (
\begin{array}{ccc}
1 &-\epsilon'_1/ \tilde{m}_2&-\epsilon'_2\\
\epsilon'_1/\tilde{m}_2& 1&-\epsilon'_3 \\
\epsilon'_2&\epsilon'_3&1
\end{array}
\right ) ,
\]
where $\epsilon_i$ and $\epsilon'_i$ are small and determined by powers of $\epsilon$ (see below). The $\tilde{m}_i$ are the ratios of the two light mass eigenvalues to the third generation mass.

This leads to
\begin{equation}
S_{u,d}\approx \left(
\begin{array}{ccc}
1 & \frac{\epsilon^{u,d}_1}{\tilde{m}_{u,d2}} & \epsilon^{u,d}_2 \\
-\frac{\epsilon^{u,d}_1}{\tilde{m}_{u,d2}} & 1& \epsilon^{u,d}_3 \\
-\epsilon^{u,d}_2 & -\epsilon^{u,d}_3 & 1
\end{array}
\right).
\end{equation}
For the case of the $U(1)$ models (models I,II amd III), expressing $S$ in terms of the horizontal $U(1)$ charges one gets

\begin{equation}
S_{u,d}\approx \left (
\begin{array}{ccc}
1&\epsilon_{u,d}^{|q_1+d_2|-|q_2+d_2|} & \epsilon_{u,d}^{|q_{1} - q_{3}|} \\
-\epsilon_{u,d}^{|q_1+d_2|-|q_2+d_2|} &1&\epsilon_{u,d}^{|q_{2} - q_{3}|} \\
-\epsilon_{u,d}^{|q_{1} - q_{3}|}&-\epsilon_{u,d}^{|q_{2} - q_{3}|}&1
\end{array}
\label{ckm}
\right ) ,
\end{equation}
where $\epsilon_{u,d}=\theta/M_{U,D}$ and for model III the charge should be evaluated setting $\omega=0$. It is straightforward to determine S for the remaining models and the result in all have the form 
\begin{equation}
S_{u,d}\approx \left (
\begin{array}{ccc}
1&\epsilon_{u,d} & \epsilon_{u,d}^3 \\
-\epsilon_{u,d}&1&\epsilon_{u,d}^2 \\
-\epsilon_{u,d}^3&-\epsilon_{u,d}^2&1
\end{array}
\label{ckm}
\right ) .
\end{equation}
Note that the $S_{u,d}$ is determined entirely by the charges of the left-handed doublet fields. 
For the $U(1)$ models the unitary matrices needed to go to the mass basis of the singlet quarks are given by
\begin{equation}
V_{D}\approx \left (
\begin{array}{ccc}
1&\epsilon_{d}^{|q_2+d_1|-|q_2+d_2|} & \epsilon_d^{|d_{1} - d_{3}|} \\
-\epsilon_{d}^{|q_2+d_1|-|q_2+d_2|}&1&\epsilon_d^{|d_{2} - d{3}|} \\
-\epsilon_d^{|d_{1} - d_{3}|}&-\epsilon_d^{|d_{2} - d_{3}|}&1
\end{array}
\right ) 
\end{equation}
and $V_{U}$ is given by the same form with $u$ instead of $d$. It is again straightforward to determine V for the remaining models. For all the models the resulting mixing matrices are given in terms of the Yukawa couplings listed in the Appendix by 
\begin{equation}
V_{D}\approx \left (
\begin{array}{ccc}
1&Y_{D,21}/Y_{D,22} &Y_{D,22} \\
-Y_{D,21}/Y_{D,22}&1&Y_{D,32} \\
-Y_{D,31}&-Y_{D,32}&1
\end{array}
\right ) 
\end{equation}
and the equivalent form for $V_U$.

We are now ready to analyze the family symmetry implications for the magnitude of the dimension 3 two fermion flavour changing operators.
Consider first the first operator in Table 2 with down quarks as  the external quarks. In MFV $X_{LL}^Q$ is $Y_u Y^{\dagger}_u$ transformed to EWDD and, for the  $U(1)$  models considered here, is given by
\begin{equation}
X^{Q,MFV}_{LLij} = \lambda^{2}_t U_{3i}^{\dagger}U_{3j} \sim \lambda^{2}_t \epsilon^{|q_i - q_3| + | q_j- q_3|}. 
\end{equation}

Transformation to EWDD of the relevant first flavon operator of equation (\ref{dim6flavons}) gives

\begin{equation}
\bar Q_L \Phi_La^{LL} \Phi^{\dagger}_L Q_L \rightarrow \bar Q_L S^{\dagger}_d (\Phi_La^{LL} \Phi^{\dagger}_L ) S_d Q_L .
\end{equation}
so the equivalent coupling for the $U(1)$ family symmetry case is given by 
\begin{equation}
X^{Q,U(1)}_{LLij} =\left ( S^{\dagger}_d (\Phi_L a^{LL}\Phi^{\dagger}_{L}) S_d \right )_{ij}\label{X}\ . 
\label{xll}
\end{equation}

For  the first two $U(1)$ models of Appendix 1 $ q_i>0$ and thus $X^{Q,U(1)}_{LLij}  \approx \epsilon^{ |q_{Li} - q_{Lj}|} $.
For $i$ or $j$ equal to 3 the magnitude is the same for MFV and for the $U(1)$ family symmetry. However there is a difference for  $ij=12$. We have (for $\lambda_t=1$)
\begin{equation}
X^{Q,MFV}_{LL12} \propto \epsilon^{|q_1 | +| q_2 | },\;\;
X^{Q,U(1)}_{LL12} \propto \epsilon^{|q_1 -q_2| }.
\end{equation}

For the  model III the situation is different since the contribution from $S_d$ in equation (\ref{xll}) is governed by the charge $|q_1|-|q_2|= 1$ while the contribution from $\Phi_L$ is governed by the charge $|q_1-q_2|=5$. In this case $X^{Q,U(1)}_{LL12}$ is dominated by $S_{d12}$ and is of the same order as for the first two models. However if, in the absence of  family symmetry breaking, the interactions are family blind there is a GIM cancellation that eliminates this contribution from $S_d$. This is clear from equation (\ref{X}) because the family blind assumption requires $a^{LL}_{11}=a^{LL}_{22}$ and then the contribution to the $(1,2)$ matrix element cancels between the $S_d^{\dagger}$ and $S_d$ contribution. In what follows we will take the extreme  family blind  case in our estimates of the possible suppression from family symmetries  so for model III too we have $X^{Q,U(1)}_{LLij}  \approx \epsilon^{ |q_{Li} - q_{Lj}|} $. 

The analysis is readily extended to the $U(1)\times U(1)$ model IV and the results are summarised in  Table \ref{deftabc1}. Note that in this case it is not necessary to assume the family blind assumption in the down sector before spontaneous breaking because, in this model, the rotations needed to diagonalise the down quark sector are very small. However we have assumed the up sector is family blind when computing the up quark operator suppression factors given in the table. In the table the operator charges $p$  are listed; the associated operator coefficients are given by  $\epsilon^{|p| }$. A similar notation is used in the case of the F-theory models involving three Abelian factors. In this case $X^D_{LR}$ and $X^U_{LR}$ do not appear except in combination with a Higgs fields suppressed by  vevs.  We shall return to a more detailed discussion of the family blind assumption in the supersymmetric context.

In the model based on a non-Abelian symmetry it is a {\it prediction} of the symmetry that, in the absence of family symmetry breaking, the interactions are family blind and so $X^{Q,U(1)}_{LLij}$ is given entirely by the spurion contribution. The symmetries of the model \cite{Ivo:2006ma} limit the spurion combinations to $\Phi_3\Phi_3^{\dagger}$, $\Phi_{23}\Phi_{23}^{\dagger}$, $\Phi_{23}\Phi_{123}^{\dagger}$ and $\Phi_{123}\Phi_{123}^{\dagger}$ and this leads to the suppression factors shown in the fifth column of Table \ref{deftabc1}.

As a second example consider  the fourth operator in Table 2. In MFV the leading term transforming as $(3,1,\bar{3})$ is $Y^{\dagger}_d Y_u Y^{\dagger}_u$ so in the MFV the operator matrix of coefficient $X^{D,MFV}_{RL}$ is given by $Y^{\dagger}_dX^{Q,MFV}_{LLij} $. 

For the first two Abelian $U(1)$ models we have
\begin{equation} \label{DRLQ}
\bar Q_L \Phi_L a^{LD} \Phi^{\dagger}_D D_R \rightarrow \bar Q_L S^{\dagger}_d ( \Phi_L a^{LD} \Phi^{\dagger}_D ) V_d D_R \sim 
\sum_{k,p} \bar Q_{Li} \epsilon^{|q_i -q_k| + | q_k + d_p | + | d_p - d_j| } D_{Rj},
\end{equation}
so the equivalent coupling
is 
\begin{equation}
X^{D,U(1)}_{LR ij} \approx \sum_{k,p} \epsilon^{|q_i -q_k| + | q_k + d_p | + | d_p - d_j| }. 
\end{equation}

For the third model the result takes a different form due to the appearance of negative charges that change the form of $S_d$ and $V_D$.
For the non-Abelian model the structure is the same as that for the first operator considered above because the LH and charge conjugate RH states have the same transformation property under the family symmetry.

A final example is given by the second operator of Table 2. It has MFV structure
\begin{equation}
\bar D_R Y^{\dagger}_d Y_uY^{\dagger}_u Y_D D_R \rightarrow \bar D_R \lambda_d \lambda_{FC} \lambda_d D_R. 
\end{equation}
and for the $U(1)$ models is
\begin{equation}
\bar D_R ( \Phi_D a^{DD}\Phi_D^{\dagger} D_R \rightarrow \bar D_R V^{\dagger}_D ( \Phi_D a^{DD}\Phi_D^{\dagger} ) V_D D_R \sim
\bar D_{Ri} \epsilon^{| d_i - d_j| }D_{Rj} ,
\end{equation}
where, as before, we have assumed for the  Model III that, in the absence of family symmetry breaking, the interactions are family blind.

\begin{table}[hdp]
\vspace{-2cm}
\begin{center}
$$
\begin{array}{|c|ccccccc|}
\hline \hline 
X^{Q}_{LL \, ij}=\Phi_{L \, i} \otimes \Phi^{\dagger}_{L \, j} & {\rm M \; I} & {\rm M \; II} & {\rm M \; III} &{U(1)^2} &{\rm N-A}&F-$theory$ & {\rm MFV} \\
\hline \hline
(12)  & 1&1& -5&(3,-1)\sim   5&3&(-2,0,0)~5 &5\\
(13) & 3 & 3 & -3&(3,0)\sim   3& 3&(-1,1,-5)~3& 3\\
(23) & 2 & 2 & 2 &(0,1)\sim   2& 2&(1,1,-5)~2& 2\\
\hline \hline
\end{array}
$$

\end{center}

\begin{center}
$$
\begin{array}{|c|ccccccc|}
\hline \hline 
X^{D}_{RR\, ij}=\Phi_{D \, i} \otimes \Phi^{\dagger}_{D \, j} & {\rm M \; I} & {\rm M \; II} & {\rm M \; III} & U(1)^2& {\rm N-A} &F-$theory$& {\rm MFV}  \\
\hline \hline
(12)  & 1&1& -5&(-5,3)\sim  {11}&  3&(-2,0,0)~5& 5 (\lambda_d \lambda_s )\\
(13) & 1 & 1 & -5&(-1,1)\sim   3& 3&(-2,2,5)~5& 3( \lambda_d \lambda_b )  \\
(23) & 0 & 0 & 0 &(4,-2)\sim   8& 2&(0,2,5)~4&2 ( \lambda_s \lambda_b )  \\
\hline \hline
\end{array}
$$
\end{center}

\begin{center}
$$
\begin{array}{|c|ccccccc|}
\hline \hline 
X^{U}_{RR \, ij}=\Phi_{U \, i} \otimes \Phi^{\dagger}_{U \, j} & {\rm M \; I} & {\rm M \; II} & {\rm M \; III} &U(1)^2& {\rm N-A} &F-$theory$ &{\rm MFV} \\
\hline \hline
(12)  & 1&1& -5&(2,2)\sim   6& 3&(-2,0,0)~5& - \\
(13) & 3 & 3 & -5&(-1,2) \sim   5& 3&(-1,1,-5)~5& - \\
(23) & 2 & 2 & 0 &(1,0) \sim 1&2&(1,1,-5)~4& - \\
\hline \hline
\end{array}$$
\end{center}
\begin{center}
$$
\begin{array}{|c|ccccccc|}
\hline \hline 
X^{D}_{LR \, ij}=\Phi_{L \, i} \otimes \Phi^{\dagger}_{D \, j} & {\rm M \; I} & {\rm M \; II} & {\rm M \; III} &U(1)^2&  {\rm N-A} &F-$theory$& {\rm MFV} \\
\hline \hline
(12)  & 3&4& -3+w&(7,-1)\sim   9& 3&(-2,0,-3)~-&  5( \lambda_s) \\
(13) & 3 & 4 & -3+w&(3,1) \sim   5& 3& (-2,2,2)~-& 3 (\lambda_b)\\
(23) & 2 & 3 & 2+w &(0,2) \sim   4& 2 & (0,2,2)~-& 2(\lambda_b)\\
\lambda_d & 4 & 5& 4 &(2,2)\sim   6& 4 &&{} \\
\lambda_s & 2 & 3& 2 &(4,0)\sim   4& 2 &&{} \\
\lambda_b & 0 & 1& 0 & (0,1) \sim   2&0 &&{} \\
\hline \hline
\end{array}
$$
\end{center}
\begin{center}
$$
\begin{array}{|c|ccccccc|}
\hline \hline 
X^{U}_{LR \, ij}=\Phi_{L \, i} \otimes \Phi^{\dagger}_{U \, j} & {\rm M \; I} & {\rm M \; II} & {\rm M \; III} & U(1)^2& {\rm N-A} &F-$theory$& {\rm MFV} \\
\hline \hline
(12)  & 5&5& -3+w&(4,0) \sim   4& 3 &(-2,0,0)-&- \\
(13) & 3 & 3 & -3+w&(3,0)  \sim   3& 3&(-1,1,-5)-~& -\\
(23) & 2 & 2 & 2+w&(0,1) \sim   2& 2 &(0,0,0)-&-\\
\lambda_u & 6 & 6& 8 &(2,2) \sim   6&4 &&{} \\
\lambda_c & 4 & 4& 2 &(1,1) \sim   3&2 &&{}\\
\lambda_t & 0 & 0& 0 &(0,0) \sim 0&0&&{}\\
\hline \hline
\end{array}
$$
\end{center}
\caption{{\bf Charge structure of the  dimension 3 operators of Table 2.}{\it The coefficient of the operator is given by $\epsilon^{|p|} $ where $p$ is the charge. For the $U(1)^2$ model the coefficient is $\epsilon^{|p_1|+2|p_2|}$}}.

\label{deftabc1}
\end{table}

In Table \ref{deftabc1} we list the resulting charges associated with the various matrix elements of the dimension 3 operators given in Table 2. The first 5 columns give the charge structure of the operator coefficients $X$ for the models introduced in Appendix 1. The associated operator coefficients are simply given by $\epsilon^p$ where $p$ is the modulus of this charge. For  the $U(1)^2$ model we also show in parenthesis the underlying coefficient in terms of the two expansion factors. For comparison we show the equivalent charges for the MFV . In parenthesis we give the Yukawa coupling factor that must also be included when the external quarks are not LH down quarks; these are so small that the operator is usually dropped in MFV.

\section{Comparison of MFV and family symmetry models -
 dimension 6 four quark operators}
Of course in phenomenological studies it is the dimension 6 flavour changing operators of the type shown in Table 1 that are relevant. 

\subsection{Factorisation of operators}
We start with a discussion of which dimension 6 operators factorise in the sense that they are determined by the coefficients of the dimension 3 bilinear operators discussed in the last Section. Note that the factorisation applies to all operators for the case of MFV.

\subsubsection{$\Delta F_i=1,\; \Delta F_j=-1,\; i\ne j$ operators}
In our notation, $\Delta F_i=\pm 1$ means a change by one unit of the $i-th$ flavour, for instance the operator $(\bar b....s)$ annihilates a quark $s$ and creates a quark $b$, so 
$\Delta F_2=-1$, $\Delta F_3=+1$.
These operators include $O_{F1,G1,l1,l2,H1,q5} $ of Table 1 together with related operators involving up quarks. For the operators involving only two quarks it is obvious that the flavour changing component comes from the quark bilinear operator and  so the dimension 6 coefficient is determined by equivalent coefficient of the dimension 3 operator. This class of operator also involves operators involving four quarks, such as $O_{q5}$ that have family change only in one factor. 

In the Abelian family models, up to $O(1)$ factors, the  operator coefficient is determined by the overall sum of the $U(1)$ charges. For the operator $O_{q5}$ the charges of the second bilinear factor sum to zero and the coefficient is determined by the first quark bilinear operator alone. For  the operator related to $O_{q5}$ by a Fierz transformation the overall charge clearly remains the same and so its coefficient is also determined by the flavour changing quark bilinear operator formed when Fierz transforming back to the form of $O_{q5}$. The same conclusion applies to the other four quark operators of this type. 

For  the non-Abelian family symmetry the structure is somewhat different because the number of familon insertions may change for the operators related by Fierz transformations if the Fierz transformation results in two quark bilinear factors each of which involves flavour change. In this case the leading term corresponds to the ordering of the operator with flavour change in a single bilinear factor and this factor alone determines the coefficient.

\subsubsection{$\Delta F_i=2,\;\Delta F_j=-2,\;i\ne j$ operators}
An example of this class of dimension 6 four quark operator is given by the operator $O_0$. Since it involves the square of a dimension 3 two quark operator the coefficient is determined by the square of the coefficients of the quark bilinear operator. Again this factorisation is only up to $O(1)$ factors. For this class of operator Fierz transformation does not affect this structure.

\subsubsection{Non-factorisable operators} There are several types of operator that do not factorise. An example is the $\Delta F_i=2,\;\Delta F_j=-1,\;\Delta F_k=-1,\; i\ne j \ne k$ operators.  Suppressing the Lorentz structure, an example of these dimension 6 four quark operators is given by $(\bar Q_{L1}Q_{L3})(\bar Q_{L2}Q_{L3})$. Here $\Delta F_3=-2$,
$\Delta F_1=\Delta F_2=+1$. Depending on the particular form of the family symmetry the coefficients of these operators may {\it not} factorise into the product of any combination of the quark bilinear pairs that make up the operator.
Further examples of non-factorising operators are $\bar Q_{Li} U_{Rj} \bar Q_{Lk} D_{Rl} X^{ijkl}$, $\bar Q_{Li} Q_{Lj} \bar U_{Rk} U_{Rl} Y_{1}^{ijkl}$ and $\bar Q_{Li} Q_{Lj} \bar D_{Rk} D_{Rl} Y_{2}^{ijkl}$ with family change in both of the factors.

\subsection{Determination of the coefficients of the dimension 6 operators} For the $\Delta F_i=1,\; \Delta F_j=-1,\; i\ne j$ operators the dimension 6 operator coefficients are given by the coefficient associated with the appropriate flavour changing dimension 3 two quark operator. As discussed above this is determined by $\epsilon^x$ where $x$ is the modulus of the associated charge listed in Table \ref{deftabc1}.  One exception to this rule are the coefficients of the operators $\cO_{F1}$ and $\cO_{G1}$  in the Model III where  the horizontal charge $-\omega$  of the Higgs field has to be taken into account. For the Abelian family symmetries these coefficients are  determined up to an $O(1)$ factor but in the case of the non-Abelian family symmetry the relative magnitude of the coefficients at a given power of $\epsilon^x$ are determined. 
The factorisable $\Delta F_i=2,\;\Delta F_j=-1,\;\Delta F_k=-1,\; i\ne j \ne k$  operator coefficient is given by the product of the appropriate flavour changing dimension 3 two quark operator.

\begin{table}[h]
$$
\begin{array}{rc|cccccccc}
\multicolumn{2}{c}{\hbox{Flavour violating}}   &\multicolumn{1}{c}{}&\multicolumn{1}{c}{}&\multicolumn{1}{c}{\Lambda/\Lambda_{MFV}}&\multicolumn{1}{c}{}&\multicolumn{1}{c}{}&\\
\multicolumn{2}{c}{\hbox{dimension~six~operator}}   & {\hbox{ Ex. 1}}&\multicolumn{1}{c}{\hbox{ Ex. 2}}&\multicolumn{1}{c}{\hbox{ Ex. 3}}&\multicolumn{1}{c}{\hbox{ $U(1)^2$}}&\multicolumn{1}{c}{\hbox{ N-A}}&F&\\

\hline
\cO_{0}= &\frac{1}{2} (\bar Q_L �X^{Q}_{LL} Q_L)^2   
\phantom{X^{X^X}}
& \multicolumn{1}{c}{  \epsilon^{-4}}&\multicolumn{1}{c}{  \epsilon^{-4}}&\multicolumn{1}{c}{  1}&\multicolumn{1}{c}{  1}&\multicolumn{1}{c}{  \epsilon^{-2}}&1&\\
\cO_{F1}= & ��H^\dagger \left( {\bar D}_R �X^{D \dagger}_{LR} \sigma_{\mu\nu}
Q_L \right) F_{\mu\nu} �   & \multicolumn{1}{c}{ x \epsilon^{-2}}&\multicolumn{1}{c}{ x \epsilon^{-3/2}}&  x\epsilon^{-2}&\multicolumn{1}{c}{  x\epsilon}&\multicolumn{1}{c}{ x \epsilon^{-2}}&\multicolumn{1}{c}{ x \epsilon^{-2}}&  \\
\cO_{G1}= & �H^\dagger \left( {\bar D}_R �X^{D \dagger}_{LR} \sigma_{\mu\nu}
�T^a �Q_L \right) �G^a_{\mu\nu} �   &  \multicolumn{1}{c}{ x \epsilon^{-2}}&\multicolumn{1}{c}{  x\epsilon^{-3/2}}& x\epsilon^{-2}&\multicolumn{1}{c}{x \epsilon}&\multicolumn{1}{c}{ x \epsilon^{-2}}&\multicolumn{1}{c}{ x \epsilon^{-2}}&  \\
\cO_{\ell1}=& (\bar Q_L X^{Q}_{LL} \gamma_{\mu} ��Q_L)(\bar L_L \gamma_\mu L_L ) �  
&\multicolumn{1}{c}{  \epsilon^{-2}}&\multicolumn{1}{c}{  \epsilon^{-2}}&\multicolumn{1}{c}{  1}&\multicolumn{1}{c}{  1}&\multicolumn{1}{c}{  \epsilon^{-1}}&1& \\
\cO_{\ell2}= &( {\bar Q}_L X^{Q}_{LL}  \gamma_\mu �\tau^a Q_L)
({\bar L}_L \gamma_\mu �\tau^a L_L)\quad ��   &\multicolumn{1}{c}{  \epsilon^{-2}}&\multicolumn{1}{c}{  \epsilon^{-2}}&\multicolumn{1}{c}{  1}&\multicolumn{1}{c}{  1}&\multicolumn{1}{c}{  \epsilon^{-1}}&1& \\
\cO_{H1}=& (\bar Q_L X^{Q}_{LL} \gamma_{\mu} Q_L)(H^\dagger i D_\mu H)\qquad �  
& \multicolumn{1}{c}{  \epsilon^{-2}}&\multicolumn{1}{c}{  \epsilon^{-2}}&\multicolumn{1}{c}{  1}&\multicolumn{1}{c}{  1}&\multicolumn{1}{c}{  \epsilon^{-1}}& 1&\\
\cO_{q5}=& (\bar Q_L X^{Q}_{LL} \gamma_{\mu} Q_L)(\bar D_R \gamma_\mu D_R ) �\qquad &  \multicolumn{1}{c}{  \epsilon^{-2}}&\multicolumn{1}{c}{  \epsilon^{-2}}&\multicolumn{1}{c}{  1}&\multicolumn{1}{c}{  1}&\multicolumn{1}{c}{  \epsilon^{-1}}&1& \\
\end{array}$$
\label{nnnnewtable}
\caption{\em
{\bf Bounds  on the suppression scale of the familon induced operators}.
The SM is extended by adding flavour-violating
dimension-six operators with coefficient $1/\Lambda^2$. Here
we report the bounds on $\Lambda$ for the family symmetry models in terms of the bounds on $\Lambda_{MFV}$ for MFV given in Table 1. Here $x=(m_t/m_b)^{1/2}$. The bounds come from  the flavour changing operators involving  the first two families. 
}
\end{table}

Using this the resulting bounds on the scale of new physics coming from the operators listed in Table 1 are shown in Table 4 for the models of Appendix 1 relative to the MFV value given in Table 1. Note that these bounds come from the operators involving the down and strange quarks that are dominant in the MFV case. Since $x\epsilon \approx 1$ it may be seen that  all models  except the $U(1)\times U(1)$ model require a  larger mediator suppression scale to keep the FCNC associated with the operators $O_{F1}$ and $O_{G1}$ within present bounds. The reason for this is that only the $U(1)\times U(1)$ model has, in the current quark basis,  very small mixing between the first two families in the down quark mass matrix, the Cabibbo angle being generated from the mixing in the up quark sector.  

The physical interpretation of the mediator suppression scale depends on the microscopic physics that has been integrated out. In particular in supersymmetric models it {\it may} be related to the supersymmetry breaking scale and  in some cases the bounds on FCNC may be difficult to reconcile with  SUSY solving the hierarchy problem. In the next Section we shall discuss the identification of the mediator scale for the case of the Minimal Supersymmetric Standard Model (the MSSM) and in Section 7 consider the FCNC tests in SUSY models in more detail.

As noted above  the $U(1)\times U(1)$ model illustrates the fact that  family symmetry models can give approximately the  same expectation for the Table 1 operator coefficients as MFV. In this case one must turn to the other possible operators involving the third generation to distinguish them.  
We  emphasised above that, in contrast to the MFV case,  the operators appearing in Table 1 may not be the only ones contributing significantly to flavour changing processes in  the family symmetry models. For the factorising operators it is easy to use Table \ref{deftabc1}  to determine the coefficients of the remaining operators. For example for flavour changing involving the light quarks, the $(1,2)$ sector, the first three dimension 3 operators of Table 2 all have the same order of coefficients for the family models considered. This is to be compared to MFV in which only the first operator is significant {c.f. Table 1}. The second and third operators have a different Lorentz structure and consequently the implications for the phenomenological importance of the dimension 6 operators involving them may be significantly different from those involving the first operator of Table 2. It is beyond the scope of this paper to perform a complete analysis of the FCNC effects following from these terms. However in Section 7 we will consider the phenomenological implications of all the operators of Table 2 for the case of supersymmetric models.

\begin{table}[htdp]
\begin{center}
$$
\begin{array}{|c|c|c|c|c|c|c|c|c|}
\hline \hline 
\phantom{X^{X^X}}& {\rm Component} & {\rm M \, I} & {\rm M \, II}&{\rm M \, III} & U(1)^2 & {\rm N-A}&F\\ \hline \hline
1. & X^{1212} &8 &9 &|6-2w|&(11,1) \sim 13&6&5\\
2. & X^{2112} &8 &9 &|6-2w|&(6,2)\sim 10&6&5\\
3. & X^{3223} &4 &5 &|2+w|&(1,2) \sim 5&6&7\\
4. & X^{2131} &6 &7 &|8-w|&(2,5) \sim 12&6&3\\
\hline \hline
\end{array}
$$
\end{center}
\caption{\em{\bf Coefficients $X^{ijkl}$ of dimension 6 four-fermion operators of the form $\bar Q_{Li} U_{Rj} \bar Q_{Lk} D_{Rl} $. }The coefficient of the operator is given by $\epsilon^p$ where $p$ is the modulus of the charge.}
\label{deftabc6}
\end{table}%

\begin{table}[htdp]
\begin{center}
$$
\begin{array}{|c|c|c|c|c|c|c|c|c|}
\hline \hline 
\phantom{X^{X^X}}& {\rm Component} & {\rm M \, I} & {\rm M \, II}&{\rm M \, III}& U(1)^2 & {\rm N-A} &F\\ \hline \hline
1. & Y_2^{1212} & 2 & 2&10&(6,2)\sim 10&8&10\\
2. & Y_2^{1213} & 4&4&8&(6,1) \sim8&8&10\\
3. & Y_2^{1231} & 2& 2& 2&(0,1) \sim2&6&6\\
4. & Y_2^{2131} & 2& 2& 3&(6,1) \sim 8&8&10\\
\hline \hline
\end{array}
$$
\end{center}
\caption{\em{\bf Coefficients $Y_2^{ijkl}$ of dimension 6 four-fermion operators of the form $  \bar{Q}_{L,i} Q_{L,j}\bar{D}_{R,k}  D_{R,l}$.} The coefficient of the operator is given by $\epsilon^p$ where $p$ is the modulus of the charge.}
\label{deftabc7}
\end{table}%

Finally we turn to the non-factorising operators of the form $\Delta F_i=2,\;\Delta F_j=-1,\;\Delta F_k=-1,\; i\ne j \ne k$. There are many possible operators of this type because one can combine the different dimension 3 bilinear operators in many ways. In Tables \ref{deftabc6}  and \ref{deftabc7} we  illustrate the family symmetry prediction for the coefficients of these operators by just two examples. For the Abelian family symmetries the coefficient of the dimension 6 four quark operator is given by the factor $\epsilon^p$ where $p$ is the modulus of the overall charge of the operator. For the non-Abelian symmetry the coefficient is determined by identifying the product of familon fields needed for a given operator, chosen from the allowed set listed above. One sees a very wide range of coefficients and low suppression in many cases. Moreover the predicted coefficients differ significantly between models so the the observation of a specific pattern of flavour changing processes would provide strong evidence for a particular family symmetry.

\section{SUSY}
The analysis has so far considered the effective field theory relevant at energy scales below the mass of the new states responsible for generating the flavour changing operators. It is important to stress that the analysis is quite general and covers all possibilities for Beyond the Standard Model physics. 
However, as discussed above, the interpretation of the meaning of the inverse mass scale characterising the bound on the operator requires a discussion of the underlying physics origin. In this Section we discuss the case that the hierarchy problem is solved by low-energy supersymmetry but allow the flavour symmetry breaking scale to be much higher. 

\subsection{Identification of the scale $\Lambda$}
Since there are two fundamental scales it is necessary to determine the scale, or combination of scales, that is relevant to the bound on the scale, $\Lambda$, of Table 1. To answer this it is necessary to consider the leading flavour changing operators in the supersymmetric theory ${\it above}$ the supersymmetry breaking scale, $M_{SUSY}$, but below the flavour symmetry breaking scale, $M$. Since the quarks and leptons have scalar partners there are new operators that may violate flavour and the leading ones have a ${\it lower}$ dimension than the dimension 6 operators built of SM states alone. The SUSY operators generate the SM dimension 6 operators but, as we shall discuss, $\Lambda$ should not be interpreted as the flavour changing scale if the underlying SUSY operators have dimension $<$ 6. The leading SUSY operators are the soft supersymmetry breaking operators, the dimension 2 operators bilinear in the squark fields, the dimension 3 operators trilinear in the squark and Higgs fields and the dimension 4 four squark operators. The first two contribute to the Standard Model dimension 6 operators  at one loop order while the latter contributes at two loop order and is sub-leading. In general they are not diagonalised by the same rotations that diagonalise the fermion masses and in this case will induce flavour changing processes. At higher order there are SUSY operators of dimension 5 and above that are suppressed by additional inverse powers of $M$ that may also induce flavour changing processes. 

The Lagrangian involving the dimension 2 bilinear operators has the form  \\$M_{SUSY}^2 b_{ij}\phi_i^{\dagger}\phi_j$, where $\phi_{i,j}$ are both left-handed or right-handed squark fields and $i,j$ are family indices, corresponding to the squark mass matrix. In the case there is an unbroken family symmetry, both the squark and quark mass matrices are simultaneously diagonalised. However once the family symmetry is broken this is no longer the case and family symmetry breaking squark mass terms of the form $M_{SUSY}^2 \phi_i^{\dagger}\phi_j(\theta/M)^{{q}_i-{q}_j}$ are generated. The important point to note is that the family symmetry breaking scale only appears in the ratio $\epsilon=(\theta/M)$, the parameter that orders the family symmetry breaking terms. The SUSY operators subsequently generate the SM dimension 6 operators at one-loop order principally through gaugino interactions involving gauginos with mass scale $M_{SUSY}$. Thus it is the SUSY breaking scale in the visible sector and {\it not} the family symmetry breaking scale that appears in the denominator after integrating out the supersymmetric states. In this case $\Lambda=M_{SUSY}/{\alpha}$ where $\alpha$ is the one loop factor associated with the gaugino dressing - the strong fine structure constant. 

The discussion extends readily to the remaining operators. The dimension 3 terms involving LH- and RH-squarks and a Higgs scalar have a coefficient of $O(M_{SUSY})$. Thus again for them we have  $\Lambda=M_{SUSY}/{\alpha}$.  The dimension 4 operators have $\Lambda=M_{SUSY}/\alpha^2$.  There may be dimension 5 terms in the superpotential such as $Qu^c Qd^c/M$ with a single inverse power of $M$. For them the relevant scale is  $\Lambda=\sqrt{ M_{SUSY} M/ \alpha }$. 

\subsection{SUSY GIM suppression} There is a further important effect that must be taken into account when determining FCNC in supersymmetric theories, namely the supersymmetric analogue of the SM GIM mechanism that leads to a suppression of FCNC.  To discuss the contribution of the squark bilinear operators 
to the fermionic dimension 6 operators of Table 1 we have to go to the EWDD basis for fermions. For the squarks we still have a choice. A frequently used approach is to apply the EWDD rotations to supermultiplets and to work with non-diagonal
squark mass matrices. Another possibility is to go to the squark mass
eigenstate basis (by independent rotations of the fermion and scalar components of the supermultiplets), with the physics of the flavour violation by the squark sector encoded in the quark-squark-gaugino
couplings and closely resembling the GIM mechanism of the SM. To emphasize this aspect, we first discuss the latter approach for the simplified case of two generations and later we will work in the EWDD basis for supermultiplets, to make
easy use of the results already existing in the literature.

In addition to the suppression factor $\Lambda^{-2}$ the dimension 6 quark operators have a further suppression due to the SUSY GIM mechanism as we now discuss. We denote the physical squark masses by $\tilde m_i$, their squared mass difference by
$\Delta \tilde{m}^2$ and the average squark mass squared  $\tilde m^2$. Let us concentrate on the $LL$ squark mass matrices and restrict ourselves to the 2-family case. For the supersymmetry induced 1-loop coefficient to the operator $O_0$ in Table 1 one obtains the well known result 

\begin{equation} \label{rotation1}
\frac{\alpha_{s}^2}{\tilde{m}^2} \left | \sum_i \tilde{U}^{d}_{di} \tilde{U}^{d \dagger}_{is} \frac{\Delta \tilde{m}^{2}_i}{\tilde{m}^2} \right |^2 + \mathcal{O} (\frac{1}{k^2 - \tilde{m}^2})^5 , 
\end{equation}
where the elements of the matrix $\tilde{U}$ enter into the quark-squark-gluino couplings. This matrix is in general a composition of two rotations: the first is the rotation which diagonalizes the down quark mass matrix (from the original electroweak basis to the EWDD basis for quarks) and the second rotation diagonalizes the squark mass matrix (written in the original electroweak basis). Equivalently, we may look at the matrix $\tilde U$ as the one that diagonalizes the squark mass matrix transformed to the EWDD basis by the rotations on the supermultiplets.
Denoting the rotation angle in $\tilde{U}$ by $\rho$, equation (\ref{rotation1}) takes the form
\begin{equation} \label{rotation2}
\frac{\alpha_{s}^2}{\tilde{m}^2} \cos^2 (\rho) \sin^2 (\rho) \left | \frac{\tilde{m}^{2}_d - \tilde{m}^{2}_s}{\tilde{m}^2} \right |^2 .
\end{equation}
The supersymmetric GIM mechanism is evident in this formula.
Let us first consider
two limiting cases. Suppose that in the original electroweak basis the squark mass matrix is diagonal with split eigenvalues. The matrix $\tilde U$ is then given by the matrix $S_d$ of equation (13) and the angle $\rho$ is just the quark mixing angle.
In this case the effective scale $\Lambda$ associated with these operators should be identified with $\tilde{m}/(\alpha_s \sin(\rho)\Delta \tilde{m}^2/\tilde{m}^2)^n$ where $\alpha_s$ is the one loop factor and $n=1$ for the operator $O_0$ and $n=1/2$ for the other operators from the Table 1. It is dominated by the gluino contribution in which case $\alpha_s$ is the strong coupling fine structure constant divided by a numerical factor of the order of 100. To a good approximation the value of $\Delta \tilde{m}^2$, evaluated at the SUSY breaking scale is the same as it evaluated at the messenger scale associated with the communication of SUSY breaking from the hidden to the visible sector. However, due to family blind gaugino interactions, the mean mass $\tilde{m}^2$ is significantly increased in running to the low SUSY breaking scale. Phenomenological implications of the bounds on $\Lambda$ will be discussed in Section 7. 

The second limiting case we consider has the initial squark mass matrix with degenerate diagonal masses $m^2$ and with the off diagonal terms of the form $\tilde m^2 \phi_i^{\dagger}\phi_j(\theta/M)^{\tilde{q}_i-\tilde{q}_j}$. (The two mass parameters are not exactly equal because of the renormalization effects but we neglect this difference in the present discussion.) This matrix is diagonalized
by a rotation $\rho^{\prime}=45$ degrees and it dominates the quark mixing angle in the effective angle $\rho$ in equation (\ref{rotation2}),
$\rho\approx\rho^{\prime}$.
For this contribution $\Lambda=\tilde m/ (\sin(\rho')\cos(\rho')\alpha \Delta \tilde{m}^2/\tilde{m}^2)^n$ where $ \Delta \tilde{m}^2/\tilde{m}^2\approx (\theta/M)^{\tilde{q}_i-\tilde{q}_j}$. Similar results hold for the RR down squark mass matrix.

This analysis can be easily extended to the realistic $3\times 3$ case. The interplay of the effects due to the diagonal splitting and
to the off diagonal terms in the squark mass matrices in the original eletroweak basis may be then important if, for instance,
the squarks of the third generation are much lighter than the first two generations.

For easy reference to the results in the literature, we now repeat the above analysis in the EWDD
basis for the superfields, where the squark mass matrices remain non-diagonal. In this case, the mass insertion approximation can be used to calculate the one-loop diagrams.
Let us note 
that even if we start with the diagonal squark mass matrix, the rotation of the superfields to the EWDD basis generates off-diagonal entries, if the initial diagonal entries are split. Let's start with the $\tilde m^{2}_{dLL} $ sector. 
In the original electroweak basis
\begin{equation}
\tilde m^{2}_{dLL\, ij} \sim \tilde m^2 \epsilon^{|q_{Li}-q_{Lj}|} + \Delta_i \delta_{ij}, 
\end{equation}
where $\Delta_i =\tilde m_{ii}^2-\tilde m^2 $are the mass splittings on the diagonal. We again neglect the difference in the renormalization of the diagonal and off-diagonal terms (to be discussed later). The rotation of the superfields to the EWDD basis gives
(in leading order in $\epsilon$)

\begin{equation}
\left ( S^{\dagger}_{d} \tilde m^{2}_{dLL} S_d \right )_{ij} \sim \tilde m^2 \epsilon^{|q_i-q_j|} + \Delta_i S_{dij} + \Delta_j S_{dji}.
\end{equation}
Since $S_{dij}\ge \epsilon^{|q_i-q_j|}$,  the effect of the initial diagonal splitting can be  as or even more important than the contributions of the initial off-diagonal entries in this case. This result can be used to calculate the Wilson coefficients of the operators in Table 1 by integrating out the squark and gaugino degrees of freedom at one loop in the mass insertion approximation \cite{Gabbiani:1988rb}.

\subsection{Factorisation of operators in SUSY}
As we have discussed the dominant SUSY operators have dimension 2 and dimension 3 \footnote{Unless the D=4 contributions are relatively enhanced to compensate for the additional loop factor.}. Both of these are bilinear in the squark fields and one can generate all the the SM dimension 6 flavour changing operators by dressing one or two copies of these SUSY operators. Thus for these underlying dimension 2 and dimension 3 SUSY operators the factorisation of the SM operators is always the case. This means that it is possible to translate the phenomenological bounds on the dimension 6 operators to bounds on the dimension 2 and dimension 3 operators in a model independent way \cite{Gabbiani:1988rb}. As we shall discuss in the next Section this proves to be very convenient when exploring the phenomenological implications of family symmetries in SUSY theories. 

\section{Comparison with experiment}
\subsection{Experimental bounds on the squark masses}
As we discussed earlier, in a supersymmetric theory above the supersymmetry breaking scale $M_{SUSY}$ but below the flavour symmetry breaking scale $M$ there are operators bilinear and trilinear in the scalar fields that may violate flavour. These operators are not suppressed by the scale $M$ and after integrating out the supersymmetric degrees of freedom we obtain fermionic operators of dimension 6 discussed in Section 5, suppressed by the scale $M_{SUSY}$. Comparison with experimental data puts bounds on the symmetry breaking scale $M_{SUSY}$ that may depend on the theory of flavour violation at the scale $M$ (i.e. in our case on the broken family symmetry) and on the mechanism of supersymmetry breaking. The latter dependence is an additional interesting element of these considerations. For instance, in the absence of a spontaneously broken family symmetry and in the extreme case of universal soft terms and A terms proportional to the Yukawa couplings at the high scale (CMSSM), the MFV conjecture for flavour violation in the effective SM is satisfied since at low scale universality is broken only by the renormalization effects and the bounds on $M_{SUSY}$ actually do not depend on the theory of flavour violation at the scale $M$ \footnote{The MFV conjecture is more general than the CMSSM as it admits nonuniversal soft terms at the high scale provided they are consistent with MFV. Such scenarios may however be difficult to reconcile with an underlying family symmetry.}. Flavour physics in the CMSSM has been extensively studied in the literature . 

A contrasting picture emerges in gravity mediation scenarios for supersymmetry breaking with the flavour pattern of the soft terms at high scale determined solely by the broken horizontal symmetries responsible for the hierarchies in the fermion mass matrices. In this case the MFV conjecture does not apply and the bounds on $M_{SUSY}$ do depend on the underlying family symmetry.

In the following we will discuss 
the bounds on $M_{SUSY}$ in this case for the family symmetry models discussed above, using the analysis of \cite{Gabbiani:1988rb}. 
In this approach the effective fermionic lagrangian
(dimension 6 operators) is obtained by integrating out supersymmetric degrees of freedom at one loop in the EWDD basis for the chiral multiplets, that is with diagonal down quark masses but with non-diagonal squark mass matrices, with arbitrary off diagonal mass insertions. The Wilson coefficients of the fermionic dimension 6 operators depend on the dimensionful couplings of the operators bilinear and trilinear in the scalar fields, that is on the diagonal and off diagonal entries in the LL and RR blocks of the squark mass matrices and on the A-terms contributing to the LR blocks. Using phenomenological constraints (requiring that the supersymmetric contribution do not exceed the SM one-justified by the FCNC data) one obtains bounds on the ratio of off-diagonal squark mass squared insertions to the average of the diagonal mass squared terms.
Since in the family symmetry models we can calculate the off diagonal terms, the phenomenological bounds can be translated into the bounds on the diagonal entries, that is on the soft supersymmetry breaking scale in the squark sector. 
The structure of the effective lagrangian obtained in \cite{Gabbiani:1988rb} is as follows:
\begin{eqnarray}
L_{eff}&=&\frac{\alpha_s^2}{216\tilde m^2_{qij}}((\delta^d_{12\,LL})^2(\bar d_L\gamma_{\mu}s_L\bar d_L\gamma_{\mu}s_L)\times f(x) \nonumber \\
&+& (\delta^d_{12\,RR})^2(\bar d_R\gamma_{\mu}s_R\bar d_R\gamma_{\mu}s_R)\times f'(x) \nonumber \\&+& (\delta^d_{12\,LL})(\delta^d_{12\,RR})(\bar d_Rs_L\bar d_Ls_R)\times f''(x) +...+{\rm h.c.}) 
\end{eqnarray}
where $\delta_{ij \,MM}=\frac{\Delta \tilde m_{ijMM}^2}{\tilde m^2_{qij}}$,
 $\Delta \tilde m^2_{ijMM}$ , $M=L,R$, are the off-diagonal  entries in the down squark mass squared matrices.  and $\tilde m^2_{\tilde qij}=\sqrt{\tilde m^2_i \tilde m^2_j}$ is the average diagonal mass squared for the $i,j$ sector.  The loop functions
$f(x)$ where $x= \tilde m^2_{\tilde g}/\tilde m^2_{\tilde qij}$ are explicitly given in \cite{Gabbiani:1988rb} and effectively are of the order of 100. All squark masses, the gluino mass and the ratio $x$ are taken at the soft supersymmetry breaking scale $\tilde m$. 
Here we show only a few terms of the long effective lagrangian, the ones depending on   the LL and RR off- diagonal blocks in the down squark mass matrix and contributing to $\Delta F_i=2$, $\Delta_j=-2$, $i=1,j=2$ processes. For the full effective lagrangian for these
processes as well as for the processes corresponding to other values of $i,j$ and for $\Delta F_i=1, \Delta F_j=-1$
we refer the reader to \cite{Gabbiani:1988rb} (the terms describing e.g. the $B_s\bar B_s$ are missing there but the generalization is obvious).

In \cite{Gabbiani:1988rb}, model independent phenomenological bounds on various $\delta$'s are reported as a function of the average squark masses, taking account of the GIM cancellation discussed above. An updated version of the bounds can be found in \cite{Isidori:2010gz} (based on the results of \cite{Masiero:2005ua}, \cite{Ciuchini:2007cw} and
   \cite{Buchalla:2008jp})
and for LL and RR insertions and their product $\sqrt {LL\times RR}$ is given in the Table 7 for an average squark mass of $350$ GeV  and for $x=1$. For other values of the squark masses the bounds scale as $(m_{\tilde q}/350)$. The dependence of the bounds on the low energy value of the ratio $x$ is weak; they are slightly weaker for larger values of the ratio of the  gluino to squark masses. In the table we also express the $\delta$'s in terms of the expansion parameter  $\epsilon$. Up to $O(1)$ factors it  is approximately equal to the Cabbibo angle, with a range between 0.15 and 0.23. In the Table we use the lower value as a conservative estimate.

\begin{table}[t]
\begin{center}
\begin{tabular}{|cc|cc|} \hline\hline
\rule{0pt}{1.2em}%
$q$\ & $ij\ $\ &  $(\delta^{q}_{ij})_{MM}$ &
$\langle\delta^q_{ij}\rangle$ \cr \hline 
$d$ & $12$\ & $\ 0.01 \sim\epsilon^2\ $ & $\ 0.0007\ \sim \epsilon^4$ \cr
$d$ & $13$\ & $\ 0.07\sim\epsilon\ $ & $\ 0.025\ \sim\epsilon^2$ \cr
$d$ & $23$\ & $\ 0.21\sim\epsilon\ $ & $\ 0.07\ \sim\epsilon$ \cr
$u$ & $12$\ & $\ 0.035\sim\epsilon^2\ $ & $\ 0.003\sim\epsilon^3\ $ \cr
\hline\hline
\end{tabular}
\caption{\em{\bf The phenomenological upper bounds on $(\delta_{ij}^{q})_{MM}$ and
   on $\langle\delta^q_{ij}\rangle$, where $q=u,d$ and $M=L,R$ taken from the summary of  Isidori et al. \cite{Isidori:2010gz}.}
   The constraints are given for $m_{\tilde qij}=350$ GeV and $x\equiv m_{\tilde
   g}^2/m_{\tilde q}^2=1$.The masses are taken at the soft supersymmetry breaking scale.  It is assumed that the phases could suppress the
   imaginary parts by a factor $\sim0.3$. The bound on
   $(\delta^{d}_{23})_{RR}$ is about 3 times weaker than that on
   $(\delta^{d}_{23})_{LL}$ (given in table). The constraints on
   $(\delta^{d}_{12,13})_{MM}$, $(\delta^{u}_{12})_{MM}$ and
   $(\delta^{d}_{23})_{MM}$ are based on, respectively,
   Refs. \cite{Masiero:2005ua}, \cite{Ciuchini:2007cw} and
   \cite{Buchalla:2008jp}.}
 \end{center}
\label{deltaij}
\end{table} 

\subsection{Family symmetry prediction for soft masses}
\subsubsection{Contribution from non-degeneracy of squark masses and D-terms}
As discussed above, one source of the flavour changing $\phi^{\dagger}_i\phi_j$ terms arises if the squarks are not degenerate for then, in going to the appropriate quark mass eigenstate basis, off diagonal terms are generated. In the case of a non-Abelian family symmetry such as $SU(3)$ the symmetry does require that the squarks of a given flavour be degenerate. In general this is not the case but it may happen that the origin of supersymmetry breaking in the visible sector is family blind and in this case the squarks will be degenerate. This happens in gauge mediated supersymmetry breaking models and also in particular supergravity mediated models. However, even if this is the case, there is a significant additional source of non-degeneracy in models with a family gauge symmetry. This comes from the D-terms associated with the gauge symmetry. For the case of a $U(1)$ symmetry the D-term is
\begin{equation}
D^2=g_f^2 \left(|\phi|^2-|\bar\phi|^2+c_{\tilde{d}L}|\tilde{d}_L|^2 +c_{\tilde{d}R}|\tilde{d}_R|^2+... \right)^2
\end{equation}
where $g_f$ is the gauge coupling constant, $\phi$ is the familon field,  $c_{\tilde{d}L,R}$ are the family charges of the down squarks and the (...) stands for similar terms for all the other sfermions. This term gives contributions to the squark masses of the form
\begin{equation}
\Delta m^2_{\tilde{f}L,R}=c_{\tilde{d}L,R}g_f<D>
\end{equation} 
where 
\begin{equation}
<D>=g_f<|\phi|^2-|\bar\phi|^2>
\end{equation}
Following from this one has
\begin{equation}
\delta^d_{12LL}\approx \frac{<D>}{\tilde m^2}\left(c_{\tilde{d}_L}S_{d11}S^*_{d21}+c_{\tilde{s}_L}S_{d12}S^*_{d22}+c_{\tilde{b}_L}S_{d13}S^*_{d23}\right)
\end{equation}
where $\tilde m^2$ is the average squark mass squared. Similar expressions are obtained for the other $\delta$s.
As discussed in \cite{Kawamura:1994ys, Dudas:1996fe, Ivo:2006ma} the magnitude of the D-term is proportional to $(m_\phi^2-m_{\bar\phi}^2)$ where $m_i^2$ are the soft supersymmetry breaking masses squared of the familon fields. If this factor is of order $\tilde m^2$ one sees that the expectation is that $\delta^d_{12LL}$ is of order $\epsilon$.
In Table 7 we see that, for $m_{\tilde qij}=350\; GeV$ the phenomenological upper bounds on the LL and RR  $\delta$'s are at most of the order of $\epsilon^2$, and   the product $\sqrt{LL\times RR}$ in the (1,2) is bounded by $\epsilon^4$
Thus, at the first sight the D-term contribution is off by a factor $\epsilon^3$  compared to  the experimental bounds found assuming $m_{\tilde{q}ij}=350GeV$. However, these predictions are valid at the scale M of the family symmetry breaking and before comparing them to the experimental bounds one should correct them using the renormalisation group running to determine them at low scales where the experimental bounds apply. The dominant renormalisation effects are flavour blind strong interaction contributions to the diagonal squark mass entries coming from terms proportional to the gluino mass. These effects depend strongly on the ratio $x_0=m^2_{1/2} / m^2_0 $  where $m_{1/2}$ and $m_0$ are the gluino and squark masses  at the scale M \cite{Choudhury:1994pn}. For a rough estimate of such effects in the running down from the GUT scale one can use approximate formulae $m_{\tilde g}\approx 3m_{1/2}$  and $m^2_{\tilde q}\approx m^2_0+6m^2_{1/2}$.
First, we see that $x=1$ implies $x_0=1/3$ and very weak gluino renormalisation effects. The squark mass of 350 GeV corresponds then to $m_0=200$GeV and $m_{1/2}=120$ GeV. Next, we can ask for what values of $x_0$ we can gain at least
factor $\epsilon^3$ , to make the predictions consistent with the experimental bound. Neglecting the small renormalisation of the off-diagonal entries,  one finds consistency for $m_{1/2}/m_0=7$.  For 350 GeV squarks this implies $m_0=20$ GeV and $m_{1/2}=140$ GeV. For this value of $x_0$ larger values of
$m_{1/2}$ are also comfortable. For instance, for $m_{1/2}=300$ Gev we get $m_{\tilde g}=900$GeV and $m_{\tilde q}=800$
GeV, consistent with low fine-tuning \cite{Cassel:2009cx}. As a final example,
for $x_0\approx 1$ the values of the  $\delta$'s are renormalised in the running down from the GUT scale to 1 TeV by a factor of order $0.1$ and 
to bring the result into agreement with the bounds requires the squarks of the first two generations of about 15TeV. Such a large mass introduces a large fine tuning implying that SUSY does not solve the little hierarchy problem. This discussion
nicely illustrates the interplay between the FCNC effects and the soft supersymmetry breaking spectrum in models with family symmetries  \cite{Chankowski:2005jh, Ivo:2006ma}.

The magnitude of the D-term can be much smaller also for other reasons \cite{Ivo:2006ma}. One possibility in supergravity mediated SUSY breaking occurs in family symmetry models such as model III  with conjugate pairs of familons $\phi$ and $\bar\phi$. In this case the factor $(m_\phi^2-m_{\bar\phi}^2)$ vanishes for degenerate familons eliminating the D-term contribution. Such degeneracy can result if the underlying SUSY breaking field is dominantly the dilaton that couples universally.
For the case of gauge mediated supersymmetry breaking the soft familon masses are automatically much smaller that the soft squark masses because they are SM gauge singlets and their coupling to the gauge mediation sector is via their coupling to the quark, introducing an additional loop factor in the mass squared calculation. Such a factor is expected to render this contribution subdominant.

Finally it may be that the family symmetries are discrete rather than continuous and in this case there is no D-term to worry about. Although our discussion has been in the context of continuous symmetries they may also apply to their discrete subgroups. To be specific the results are unchanged for the $Z_N$ subgroup of $U(1)$ provided the operator charges are not greater than $N/2$ giving a lower bound on $N$.

\subsubsection{Contributions from off diagonal squark mass}
 Consider the bounds coming from the $LL$ terms. For them the squark mass terms in the Lagrangian have the form  $m_{\tilde q ij}\tilde q_i^{\dagger}\tilde q_j \epsilon^{|q_j-q_i|}$ corresponding to $(\delta^{q}_{ij})_{LL}=\epsilon^{|q_j-q_i|}$. This is the suppression associated with the dimension 3, $\Phi_{Li} \otimes \Phi_{Lj}^{\dagger}$, operators listed in Table \ref{deftabc1}. The other entries of Table \ref{deftabc1} immediately give the remaining suppression factors associated with the other $(\delta^{q}_{ij})_{MM}$. As for the D-terms,
these predictions are valid at the scale M of the family symmetry breaking and one should correct them using the renormalisation group running to determine them at low scales where the experimental bounds apply. The previous discussion remains valid in this case, too. Thus, the coefficients taken from Table \ref{deftabc1}  should be rescaled by a factor depending on the value of the ratio $x_0$ of the soft masses at the scale M,  before comparing them with the phenomenological bounds of Table \ref{deltaij}.

As mentioned earlier, for $m_{\tilde qij}=350\; GeV$ the phenomenological upper bounds on  the product $\sqrt{LL\times RR}$ in the (1,2) sector is $\epsilon^4$.  A comparison of Table \ref{deftabc1} with Table 7 shows that only this  term requires special attention in  some of the models for an average squark mass of $350 GeV$. 
In Models I and II, the suppression factor for the $\sqrt{LL\times RR}$ in the  (1,2) sector is only $\epsilon $ so we need
either large enough value of the ratio $x_0$ or heavier squarks or both, as discussed in the previous subsection.
The (1,3) and (2,3) sectors are still safe even for light 3rd generation squarks.
Models III, $U(1)^2$,  the non-Abelian model and the F-theory model have suppression factors of $\epsilon^5$, $\epsilon^8$, $\epsilon^3$ and $\epsilon^5$ respectively for the $\sqrt{LL\times RR}$ in the  (1,2) sector. Allowing for a modest suppression due to running of $O(0.1)$, corresponding to $x_0=1$, even for a light squark sector with masses of $O(350\; GeV)$ they are safely within the present bounds. It is interesting that Model III predicts an unsuppressed RR insertion in the (2,3) sector and an improved phenomenological bound separately on this insertion would be very interesting.

To summarise, supersymmetric family symmetry models of fermion mass generically violate the MFV hypothesis. However, they offer a broad spectrum of possibilities, from being consistent in the FCNC sector with the present experimental bounds with no constraints on the soft supersymmetry breaking parameters to requiring special pattern of SUSY breaking. Various models predict ``significant'' departures from the MFV but only in a limited number of processes involving heavy quarks suggesting a systematically study of all FCNC data may reveal deviations from MFV.

\section{A terms}

\begin{table}[t]
\begin{center} 
\begin{tabular}{|cc|c|} \hline\hline  
\rule{0pt}{1.2em}%
$q$\ & $ij\ $\ &  $(\delta^{q}_{ij})_{LR}$ \cr \hline 
$d$ & $12$\ & $2\times10^{-4} \sim\epsilon^4\ $  \cr
$d$ & $13$\ & $\ 0.08\sim\epsilon\ $  \cr
$d$ & $23$\ & $\ 0.01\sim\epsilon^2\ $  \cr
$d$ & $11$\ & $\ 4.7\times10^{-6}\sim\epsilon^6\ $  \cr
$u$ & $11$\ & $\ 9.3\times10^{-6}\sim\epsilon^6\ $  \cr
$u$ & $12$\ & $\ 0.02\sim\epsilon^2\ $  \cr  
\hline\hline
\end{tabular}
\caption{\em{\bf The phenomenological upper bounds on chirality-mixing $(\delta_{ij}^{q})_{LR}$, where $q=u,d$ taken from the summary of  Isidori et al. \cite{Isidori:2010gz}.} The constraints are given for $m_{\tilde q}=1TeV$ and for $x=m_{\tilde g}/m_{\tilde q}=1$. It is assumed that the phases could suppress the imaginary parts by a factor $\sim 0.3$. The constraints on $\delta_{12,13}^{d},\;\delta_{12}^{u},\;\delta_{23}^{d}$ and $\delta_{ii}^{q}$ are based on Refs. \cite{Masiero:2005ua}, \cite{Ciuchini:2007cw}, \cite{Buchalla:2008jp} and \cite{Raidal:2008jk} respectively (with the relation between the neutron and quark EDMs as in \cite{Gabbiani:1996hi}).}
\end{center}
\label{deltaij22}
\end{table} 

We turn now to to the $A-$terms that enter in the trilinear scalar quark couplings $A^q_{ij}H_q\tilde{q}^*_{Li}\tilde{q}_{Rj}$ where $H_q$, $q=u,d$ are the $q-$type Higgs bosons and $v_q=\langle H_q\rangle$.  These terms give rise to chirality-mixing $(\delta_{ij}^{q})_{LR}=\frac{v_qA^q_{ij}|_{SCKM}}{m^2_{\tilde qij}}$ 
squark mass insertions in the SCKM basis, where $q=u,d$ and $m_{\tilde qij}$ is the average squark mass defined above. In Table \ref{deltaij22} 
we give the current bounds on these chirality mixing masses \cite{Isidori:2010gz}.  In the table we also express the $\delta$'s in terms of the expansion parameter  $\epsilon$.

To determine the implications of these bounds for the family symmetry models note that in them $A^q_{ij}$ are suppressed by the same powers of $\epsilon$ as the Yukawa couplings $Y^q_{ij}$ given in Appendix 1. In such models, $A^q_{ij}=\tilde A^q_{ij}Y^q_{ij}$ where the coefficients  $\tilde A^q_{ij}$ are  given by an overall mass scale factor multiplied by  ${\cal O} (1)$  constants.
Rotated to the appropriate basis (in the case of the operators involving d squarks the SCKM  basis and the  EWDD basis are equivalent)  $A^q_{ij}|_{SCKM} \propto \left ( S^{\dagger}_{d} A^q V_d \right )_{ij}$.
In all examples of charge assignments considered in this paper, the  off-diagonal $A^q_{ij}|_{SCKM}$ are also suppressed by the same powers of $\epsilon$ as the Yukawa couplings $Y^q_{ij}$ given in Appendix 1.  Assuming for the moment that the constant of proportionality is the average squark mass the  chirality-mixing $(\delta_{ij}^{q})_{LR}\propto Y^q_{ij}v_q/m_{\tilde qij}$.  Comparing with the factors of Appendix 1 and taking into account that $v_q/m_{\tilde qij}<\epsilon$  one sees that the bounds are satisfied in all cases in Model IV. In the other  models the bounds are satisfied for the off-diagonal entries. In model I   the $(\delta_{11}^{d})_{LR}$ entry is  too large by the factor $1/\epsilon$. 
So, for squark masses in the TeV range all but Model I satisfy the bounds, the discrepancy with Model I being in comparison with the $(\delta_{11}^{d})_{LR}$ entry. Squark masses in the TeV range are perfectly acceptable from the point of view of still solving the fine-tuning problem \cite{Cassel:2009cx}.

The strong bounds on $(\delta_{11}^{q})_{LR}$ come from the bounds on the electric dipole moments (EDM) and, being CP violating,  are sensitive to the phase of $(\delta_{11}^{q})_{LR}$.  In supersymmetric models there are new CP violating phases (beyond those associated with $(\delta_{11}^{q})_{LR}$) associated with the gaugino mass and $\mu$-term that must be less that $10^{-2}$ to be consistent with the EDMs. In family symmetry models there is a very natural explanation \cite{Ross:2004qn}  for this suppression that follows if the underlying supersymmetric theory is CP  conserving and CP  is spontaneously  broken by the familon vevs. Taking the phases of the familon vevs to be of $O(1)$ one readily generates the observed CP violation while keeping the gaugino mass and $\mu$-term phases below $10^{-2}$. In this case the leading contribution to EDMs is that coming from $(\delta_{11}^{q})_{LR}$. The bound on the latter assumes that the phases could suppress the imaginary parts by the factor 0.3 and we consider this a reasonable estimate. 

The   possibility for weakening the bounds on the $A$-terms following from the bounds on $(\delta_{ij}^{q})_{LR}$
and making them  consistent with squarks lighter than 1 TeV
is that the constant of proportionality is much less than the average squark mass scale. As discussed in \cite{Ross:2002mr} in the case of gravity mediated supersymmetry breaking the average squark mass {\it is} the normal expectation for the constant of proportionality but in particular cases this may be significantly reduced \cite{Antusch:2008jf}. In the latter case the reduction can be by a factor of $1/\epsilon$.  In gauge mediated supersymmetry breaking the situation changes dramatically because the constant of proportionality is then expected to be much less than the average squark mass. In this case the constraint on the $A-$terms coming from $(\delta_{ij}^{q})_{LR}$  go away.

\section{Summary and Conclusions}
The precise measurements of and limits on flavour changing neutral currents and CP violation provide sensitive tests of the Standard Model and strong limits on physics beyond the Standard Model. To date there is no definitive indication of a deviation from the Standard Model predictions so one obtains bounds on the effective suppression scale  of the leading 
higher dimension operators contributing to such processes. Since the most stringent of these bounds are in the hundreds of TeV range, significantly above the TeV range expected for new physics capable of solving the hierarchy problem,  the nature of the new physics must have a  mechanism  leading to a strong suppression of FCNC effects.

A minimal possibility is that all flavour changing and CP violation originates from the Yukawa couplings of the Standard Model processes and its generalisation, such as supersymmetry, that is responsible for solving the hierarchy problem. Within this framework there is no tension with the current bounds on the mediator scale and the scale of new physics needed to solve the little hierarchy problem. However MFV does not address the origin of the Yukawa couplings and it is of interest to ask whether models that do can still satisfy the bounds and, if so, how one will be able to distinguish them from MFV. In this paper we have discussed this question in the context of spontaneously broken family symmetries that are able to generate the hierarchical pattern of fermion masses and mixing angles. 

The structure of family symmetry models is significantly different from MFV. In MFV the Yukawa couplings act as fundamental spurions with definite transformation under $SU(3)^3_q$ and all FCNC operators are built using combinations of these spurions. However in the case of Abelian family symmetries one generates all possible $SU(3)^3_q$ representations from the familon fields so FCNC are not so tightly constrained. Having set up the general formalism to deal with such structures we considered a set of representative models to get an indication of the magnitude of FCNC and CP violation to be expected in family symmetry models.
 
For the case that the structure beyond the Standard Model is not specified one obtains new bounds on the effective mediator scale needed to suppress the dimension 6 FCNC quark operators. The most sensitive case turns out to be for the operator $H^\dagger \left( {\bar D}_R \lambda_d \lambda_{\rm FC} \sigma_{\mu\nu}Q_L \right) F_{\mu\nu} $ which has the same family symmetry  property as the down quark mass matrix in the current quark basis. All but one of the models considered here have $U_{us}$ originating largly from the down quark sector and for them the bound on the mediator scale is enhanced by a factor of $O(100)$ relative to that found in MFV. To avoid this it is necessary that  $U_{us}$ comes dominantly from diagonalising the up quark sector and for it the bound is the same as that found in MFV.  The $U(1)\times U(1)$ model of \cite{Leurer:1993gy} provides an example of this and  illustrates that family symmetry models of fermion mass do not necessarily require much stronger bounds on the scale of new physics than that found in MFV. Of course if deviations from the Standard Model are found it will be crucial to be able to distinguish between MFV and family symmetry models and ultimately to determine if a given family symmetry model is correct. The study presented here shows that this may be possible through the observation of correlations of FCNC effects in a wide variety of channels because the different models considered here vary greatly in their predictions for various FCNC processes involving the different families.

A particularly interesting question is whether any of these family symmetry models is consistent with the solution to the little hierarchy problem that typically requires new physics at a scale below that found for the effective mediator mass. In the case of supersymmetric models the most dangerous SUSY terms capable of generating FCNC are the SUSY breaking squark masses and the soft trilinear scalar `A' terms.  For the former the D-terms associated with continuous family symmetries are problematic and we considered them in detail. While the present bounds  on FCNC do impose strong constraints on these terms we demonstrated that there are several ways these constraints can be satisfied without reintroducing the little hierarchy problem. The same is true for the FCNC originating from the off diagonal terms in the squark mass matrices and the soft A terms. 

The conclusion is that supersymmetric models with spontaneously broken family symmetries {\it are} consistent with all present bounds on FCNC and CP violation without the need to raise the scale of squark masses beyond that needed to solve the little hierarchy problem. However there is not much room for manoeuvre and one may expect FCNC to be close to the present bounds. If they are found then there will be characteristic signals capable of distinguishing between the models and MFV coming from the study of a variety of processes involving different family combinations. For example the recent indication of CP violation beyond the Standard Model in the D system \cite{d0} may be difficult to reconcile with MFV in which the CP violation is strongly constrained as it has to come from the Yukawa couplings alone. However in family symmetry models there are more sources of CP violation possible coming from the (possible complex) familon vevs. In addition, as we have discussed above, in family symmetry models there are additional operators, such as that associated with $Y_2^{2323}$, contributing to $\Delta B=2,\;\Delta S=-2$ processes.

 \section*{Acknowledgments}

GGR and ZL would like to thank B. Grinstein and D. Ghilencea for useful discussions. ZL and SP thank P. Langacker, M. Misiak
and A. Weiler  for  useful discussions. 
SP thanks the Galileo Galilei Institute for Theoretical Physics for the hospitality and the INFN for partial support during the completion of this work.
SP thanks the Institute for Advanced Studies at TUM, Munich,
for its support and hospitality. This work is partially supported by the European Research and Training
Network  (RTN) grant `Unification in the LHC era' (PITN-GA-2009-237920), by the EC 6th Framework Programme MRTN-CT-2006-035863  `UniverseNet' and by the  MNiSZW scientific research grant   N N202 103838 (2010 - 2012).

\section*{Appendix 1:  Family symmetry models}
To illustrate the expectations for FCNC following from a family symmetry model we consider specific models that have been built to explain the quark masses and mixings.

\section*{Example I}
The first model provides an example of an $U(1)$ holomorphic model with the familon field, $\theta$, carrying only negative charge $+1$. It is Model 1 of 
\cite{Chankowski:2005qp} with charges given by: 
\begin{eqnarray}
q_{L \; 1,2,3}: & (3,2,0) & \nonumber \\
d^c_{ 1,2,3}: & (1,0,0) & \nonumber \\
u^c_{ 1,2,3}: & (3,2,0) & 
\end{eqnarray}
This gives the following Yukawa matrices, taking $\epsilon = \frac{<\theta>}{M_P}$:

\begin{center}
\[ Y_U = \left (
\begin{tabular}{ccc}
$\epsilon^6$&$\epsilon^5$&$\epsilon^3$\\
$\epsilon^5$& $\epsilon^4$& $\epsilon^2 $\\
$\epsilon^3$&$\epsilon^2$&$1$
\end{tabular}
\right )\;
Y_D = \left (
\begin{tabular}{ccc}
$\epsilon^4$&$\epsilon^3$&$\epsilon^3$\\
$\epsilon^3$& $\epsilon^2$& $\epsilon^2 $\\
$\epsilon$&$1$&$1$
\end{tabular}
\right ) \]
\end{center}





\section*{Example II}

A second $U(1)$ holomorphic example \cite{Chankowski:2005qp} has the charge assignement: 
\begin{eqnarray}
q_{L \; 1,2,3}: & (3,2,0) & \nonumber \\
d^c_{ 1,2,3}: & (2,1,1) & \nonumber \\
u^c_{1,2,3}: & (3,2,0) & 
\end{eqnarray}
This gives the following Yukawa matrices

\begin{center}
\[
Y_U = \left (
\begin{tabular}{ccc}
$\epsilon^6$&$\epsilon^5$&$\epsilon^3$\\
$\epsilon^5$& $\epsilon^4$& $\epsilon^2 $\\
$\epsilon^3$&$\epsilon^2$&$1$
\end{tabular}
\right ) \;
 Y_D = \left (
\begin{tabular}{ccc}
$\epsilon^5$&$\epsilon^4$&$\epsilon^4$\\
$\epsilon^4$& $\epsilon^3$& $\epsilon^3 $\\
$\epsilon^2$&$\epsilon$&$\epsilon$
\end{tabular}
\right ) \]

\end{center}


\bigskip

\section*{Example III } 
The third example is a non-holomorphic model that has not previously been discussed. In addition to having the good prediction for $V_{cb}=O(m_s/M_b)$ it also has a $(1,1)$ texture zero giving the relation $V_{us}=O(\sqrt{m_s/m_d})$. In this case there are two familon fields, $\theta,\bar \theta$, with charges $\pm 1$ and equal vevs to ensure D-flatness. The Higgs fields have charge $-\omega$ and the quark charges are
\begin{eqnarray}
q_{L \; 1,2,3}: & (-3+w,2+w, w) & \nonumber \\
d^c {\; 1,2,3}: & (-5,0,0) & \nonumber \\
u^c{ \; 1,2,3}: & (-5,0,0) & 
\end{eqnarray}
where $w$ is a free parameter.
It gives the following Yukawa matrices:

\begin{center}
\[
Y_{U,D} = \left (
\begin{tabular}{ccc}
$\epsilon_{u,d}^{8}$&$\epsilon_{u,d}^{3}$&$\epsilon_{u,d}^{3}$\\
$\epsilon_{u,d}^{3}$& $\epsilon_{u,d}^{2}$& $\epsilon_{u,d}^{2} $\\
$\epsilon_{u,d}^{5}$&$1$&$1$
\end{tabular}
\right ) \]

\end{center}
where $\epsilon_{u,d} = \frac{<\theta>}{M_{U,D}}$ and we have allowed for different messenger masses in the up and the down sectors.

\section*{Example IV: A $U(1)\times U(1)^\prime$ Model}

The charges are defined in Table 9, see also \cite{Leurer:1993gy}.
\begin{table}[htdp]
\begin{center}
$$
\begin{array}{|c|c|c|}
\hline \hline 
& $U(1)$& $U(1)$^{\prime} \\ \hline \hline
 \bar{Q}_1&-3&0  \\
 \bar{Q}_2&0&-1  \\
 \bar{Q}_3&0&0  \\
 D_1&1&-2  \\
 D_2&-4&1  \\
  D_3&0&-1  \\
   U_1&1&-2  \\
   U_2&-1&0  \\
   U_3&0&0  \\
\hline \hline
\end{array}.
$$
\end{center}
\label{u1xu1}
\caption{\bf Charges in the $U(1)^2$ model. 
}

\end{table}%
The expansion parameter for the $U(1)$ is $\epsilon_1$ and for the $U(1)^\prime$ it is $\epsilon_2$. We shall assume (after \cite{Leurer:1993gy}) that 
$\epsilon_1 \sim \epsilon, $ and $\epsilon_2 \sim \epsilon^2$.  The resulting mass matrices are 
\begin{center}
\[
Y_U = \left (
\begin{tabular}{ccc}
$\epsilon^6$&$\epsilon^4$&$\epsilon^3$\\
$\epsilon^7$& $\epsilon^3$& $\epsilon^2 $\\
$\epsilon^5$&$\epsilon$&$1$
\end{tabular}
\right ) \;
 Y_D = \left (
\begin{tabular}{ccc}
$\epsilon^6$&$\epsilon^9$&$\epsilon^5$\\
$\epsilon^7$& $\epsilon^4$& $\epsilon^4 $\\
$\epsilon^5$&$\epsilon^6$&$\epsilon^2$
\end{tabular}
\right ). \]

\end{center}

\section*{Example V: A Non-Abelian Model}

The family symmetry is $SU(3)$, under which the quarks transform as follows (see \cite{Ivo:2006ma}):
\begin{equation}
Q_L \sim { \bf 3}, \; D_R, \, U_R \sim {\bf \bar 3} .
\end{equation}
The familons transform as follows
\begin{equation}
\bar \Phi_3^{u,d}\sim {\bf \bar 3}, \; \bar \Phi_{23} \sim {\bf \bar 3}, \; \bar \Phi_{123} \sim {\bf \bar 3}, 
\end{equation}
expectation values of the form: 
\begin{equation}
\bar \Phi_3^{u,d}/M_{U,D} = (0,0,1) , \; \bar \Phi_{23}/M_{U,D} = (0,1,-1) \times \epsilon_{u,d}, \; \bar \Phi_{123}/M_{U,D} = (1,1,1) \times (\epsilon_{u,d})^2,
\end{equation}
where $ \epsilon_d = 0.15$, $\epsilon_u= 0.05 \sim (\epsilon_{d})^2$. 

The allowed Yukawa couplings involving these familons are restricted by additional family independent symmetries. For the $\bar L L$ and $\bar R R$ operators these symmetries require the familon fields only appear in pairs involving the same familon field. For the $LR$ terms the familon fields appear in the combinations $\phi_{123}\phi_{23},\;\phi_{23}\phi_{23}$ and $\phi_3\phi_3$ with the corresponding mass matrices given by 
\begin{center}
\[
Y_{U,D} = \left (
\begin{tabular}{ccc}
$0$&$\epsilon_{u,d}^{3}$&$\epsilon_{u,d}^{3}$\\
$\epsilon_{u,d}^{3}$& $\epsilon_{u,d}^{2}$& $\epsilon_{u,d}^{2} $\\
$\epsilon_{u,d}^{3}$&$\epsilon_{u,d}^{2}$&$1$
\end{tabular}
\right ) \]

\end{center}
where  we have allowed for different messenger masses in the up and the down sectors.


\section*{Example VI: An F-theory model}
Recently there has been considerable interest in F-theory string models and their implications for fermion masses. Such models can have Abelian family symmetries. These symmetries and the charges of the matter fields under these symmetries are strongly constrained by the underlying $E(8)$ symmetry of the associated string theory \cite{Dudas:2009hu}. To illustrate the structure that can emerge we include here an F-theory model \cite{King:2010mq} with an underlying $SU(5)$ GUT symmetry. In this model there is a $U(1)^3$ family symmetry, a subgroup of  the $SU(5)_{\perp}$ subgroup of $E(8)$ (  $SU(5)\times SU(5)_{\perp} \subset E(8)$) when a $Z_2$ monodromy is imposed.

The charges of the quarks under these symmetries are given in Table 10. Also shown are the charges of the familon fields breaking these symmetries. There are four familon fields, $\theta_{13},\;\theta_{14},\;\theta_{53},\;\theta_{54}$ and they acquire vevs of $O(\epsilon^2,\epsilon^3,\epsilon^2,\epsilon^3 )$ respectively.

\begin{table}[htdp]
\begin{center}
$$
\begin{array}{|c|c|c|c|}
\hline \hline
& $U(1)$& $U(1)$^{\prime}&$U(3)$^{\prime \prime} \\ \hline \hline
 \bar{Q}_1&-1&1 &-2 \\
 \bar{Q}_2&1&1&-2  \\
 \bar{Q}_3&0&0&3  \\
 D_1&-1&1&1  \\
 D_2&1&1&1  \\
  D_3&1&-1 &-4 \\
   U_1&-1&1&-2  \\
   U_2&1&1&-2  \\
   U_3&0&0 &3 \\
\theta_{13}&-1&-1&5 \\
\theta_{14}&1&-1&5\\
\theta_{53}&-1&-3&0\\
\theta_{54}&1&-3&0\\
\hline \hline
\end{array}.
$$
\end{center}
\label{u1xu1}
\caption{\bf Charges in the $U(1)^3$ F-theory model.
}

\end{table}%

The Yukawa couplings have the form

\begin{center}
\[
Y_U = \left (
\begin{tabular}{ccc}
$\epsilon^6$&$\epsilon^5$&$\epsilon^3$\\
$\epsilon^5$& $\epsilon^3$& $\epsilon^2 $\\
$\epsilon^3$&$\epsilon^2$&$1$
\end{tabular}
\right ) \;
 Y_D = \left (
\begin{tabular}{ccc}
$0$&$\epsilon^3$&$\epsilon^3$\\
$\epsilon^3$& $\epsilon^2$& $\epsilon^2 $\\
$0$&$0$&$1$
\end{tabular}
\right ) \]

\end{center}

\end{document}